\documentclass[reprint,prb,longbibliography,superscriptaddress]{revtex4-2}

\usepackage{hyperref}
\usepackage{amssymb, amsmath}
\usepackage[english]{babel}
\usepackage{bm}
\usepackage{graphicx}
\usepackage[utf8]{inputenc}
\usepackage{bbold}
\usepackage{multirow}

\usepackage[dvipsnames]{xcolor}

\newcommand{\commenttoggle}[1]{}

\newcommand{\sh}[1]{{\color{ForestGreen}\commenttoggle{#1}}}

\newcommand{\red}[1]{{ #1}}

\mathchardef\myminus="2D

\usepackage{letltxmacro}
\LetLtxMacro{\ORIGselectlanguage}{\selectlanguage}
\makeatletter
\DeclareRobustCommand{\selectlanguage}[1]{%
  \@ifundefined{alias@\string#1}
    {\ORIGselectlanguage{#1}}
    {\begingroup\edef\x{\endgroup
       \noexpand\ORIGselectlanguage{\@nameuse{alias@#1}}}\x}%
}
\newcommand{\definelanguagealias}[2]{%
  \@namedef{alias@#1}{#2}%
}
\newcommand{\opvec}[1]{\hat{\mathbf{#1}}}
\newcommand{\qnm}[2]{\tilde{\mathbf{#1}}_{#2}}

\makeatother

\definelanguagealias{en}{english}
\definelanguagealias{EN}{english}
\definelanguagealias{eng}{english}
\definelanguagealias{de}{ngerman}

\begin{document}
\author{Robert Fuchs}
\affiliation{Institut für Theoretische Physik, Nichtlineare Optik und
Quantenelektronik, Technische Universität Berlin, Hardenbergstr. 36, EW 7-1, 10623
Berlin, Germany}

\author{Juanjuan Ren}
\affiliation{Department of Physics, Engineering Physics, and Astronomy,
Queen's University, Kingston, Ontario K7L 3N6, Canada}

\author{Sebastian Franke}
\affiliation{Institut für Theoretische Physik, Nichtlineare Optik und
Quantenelektronik, Technische Universität Berlin, Hardenbergstr. 36, EW 7-1, 10623
Berlin, Germany}
\affiliation{Department of Physics, Engineering Physics, and Astronomy,
Queen's University, Kingston, Ontario K7L 3N6, Canada}

\author{Stephen Hughes}
\affiliation{Department of Physics, Engineering Physics, and Astronomy,
Queen's University, Kingston, Ontario K7L 3N6, Canada}

\author{Marten Richter}
\email[]{marten.richter@tu-berlin.de}
\affiliation{Institut für Theoretische Physik, Nichtlineare Optik und
Quantenelektronik, Technische Universität Berlin, Hardenbergstr. 36, EW 7-1, 10623
Berlin, Germany}

\title{Quantization of optical quasinormal modes for spatially separated cavity systems with finite retardation
}



\begin{abstract}
A multi-cavity quantization scheme is developed using quasinormal modes (QNMs) of optical cavities embedded in a homogeneous background medium for cases where retardation is significant in the inter-cavity coupling. Using quantities that can be calculated in computational optics with numerical Maxwell solvers, we extend previous QNM quantization schemes and define a quantitative measure to determine if a separate quantization of QNM cavities is justified or if a joint quantization of the system is necessary. We test this measure for the examples of two coupled one-dimensional dielectric slabs and a dimer of metal nanorods acting as QNM cavities. For sufficiently large separations, the new scheme allows for an efficient treatment of multi-cavity phenomena using parameters defined for the individual cavities. Formulating the Hamiltonian in a familiar system-bath form, the scheme connects the rigorous QNM theory and widespread phenomenological models of open cavities coupled to a shared photonic bath with parameters obtained directly from Maxwell calculations.
\end{abstract}


\date{\today}
\maketitle

\section{Introduction}
Open photonic cavities that interact with their surrounding environment are fundamental to many advances in quantum technologies, including microlasers and single photon generation \cite{meschede1985one, scully1988two, an1994microlaser, pscherer2021single, wang2017solution, rivero2021non}, spasers \cite{bergman2003surface, noginov2009demonstration, suh2012plasmonic, song2018three, hsieh2020perovskite, epstein2020far}, and quantum information applications, including quantum computation \cite{pellizzari1995decoherence, cirac1997quantum, pellizzari1997quantum, imamog1999quantum, duan2001long, blais2004cavity, benito2019optimized, borjans2020resonant, li2021quantum}.

Especially for the latter, extended structures consisting of multiple interacting but individually addressable cavities are of great interest \cite{imamog1999quantum, cirac1997quantum, pellizzari1997quantum, aspelmeyer2003long, yin2017satellite, yin2017satellite1200}. A common theoretical approach for such systems is to assume closed cavities that are quantized \textit{separately}, and then introduce phenomenologically the coupling to other cavities or a surrounding bath \cite{cirac1991two, carmichael1993quantum, imamog1999quantum, cirac1997quantum, pellizzari1997quantum}. Such approaches are intuitive \red{and can be straightforwardly applied to quantum dynamics calculations. However, they are} inherently approximate and become problematic for cases where the dissipation exceeds the perturbative regime. Furthermore, the phenomenological coupling to other systems has to be obtained from additional calculations or experimental data. 

A more rigorous and natural approach for problems in resonant optics is to instead use cavity modes with open boundary conditions to account for losses, such as quasinormal modes (QNMs) \cite{leung1994completeness, ho1998second, muljarov2011brillouin, lalanne2018light, kristensen2020modeling},  which allow for treatment of resonators and cavities with large dissipation, including non-Hermitian and off-diagonal coupling between the modes \cite{franke2019quantization, ren2022connecting}. The QNMs have complex eigenfrequencies, leading to a temporal decay of the mode, and permit a rigorous definition of coupling elements \cite{franke2019quantization, franke2020quantized}. Optical QNMs have been applied to a variety of systems with classical \cite{leung1994completeness, kristensen2020modeling}, semi-classical \cite{muljarov2011brillouin, kristensen2012generalized, lalanne2018light}, and fully quantum \cite{franke2019quantization, franke2020fluctuation, franke2020quantized} approaches, and are routinely calculated using established (including commercial) finite-element and finite-difference time-domain solvers.

Recently, QNM theory has also been applied to coupled systems of separated resonators, both for classical \cite{el2020quasinormal, ren2021quasinormal} and quantized modes \cite{franke2019quantization, ren2022connecting, franke2022quantized}. Coupled systems are treated by either calculating joint modes for the whole system, which becomes unfeasible for various complex structures, or via QNM coupled mode theory \cite{ren2021quasinormal, ren2022connecting}, which allows for an expansion of multi-cavity quantities in terms of the single-cavity modes. Previous treatments have been focused on strongly coupled resonators with small separations, often less than the QNM wavelength. In this way, propagation delays can be neglected in the coupling between the cavities, and crucially, the \red{bound} (single-cavity) QNMs can be applied across the entire scattering geometry. 

For large spatial separations, however, the complex eigenfrequencies cause the QNMs to diverge at the positions of the other cavities\red{, making the expansion of physical quantities in terms of QNMs problematic. The divergence is related to the outgoing boundary conditions and requires a careful inclusion of retardation effects \cite{abdelrahman2018completeness}. While properly normalized QNMs can expand fields outside the resonator, such expansions usually require many modes \cite{abdelrahman2018completeness}. This is not feasible for quantum dynamics calculations, where the size of the Hilbert space scales exponentially with the number of modes, so making a few-mode expansion is required. 

Alternatively, regularized cavity fields }for positions far away from the resonator are obtained via a Dyson equation \cite{ge2014quasinormal} or the field equivalence principle \cite{ren2020near}. \red{These fields give physically meaningful expressions outside the resonator and have been used in the quantization scheme presented in \cite{franke2019quantization}.}
However, \red{regularized modes are} frequency-dependent to account for outward propagation and, thus, not immediately applicable to the established approaches for coupled resonators \red{discussed above}.

\red{Following the quantization scheme from \cite{franke2019quantization}, multi-cavity systems yield} non-bosonic operators with a finite overlap between the quantized modes. A symmetrization is performed to obtain modes with bosonic creation and annihilation operators. This involves calculating overlap integrals over the whole system, which is expensive for large structures. Multiple resonators are then described with joint sets of operators and not as individual, interacting systems as in the phenomenological models. \red{In the quantization process, information about causality is removed from modes of the system. However, the quantization scheme yields bound QNMs coupled to a photonic bath \cite{franke2020quantized}, which describes the causality removed from the modes. Thus, quantum dynamics calculations yield the full dynamics, including propagation of states between separated systems \cite{franke2020quantized, fuchs2023hierarchical}.} 

\red{In this paper, }we present a multi-cavity extension of the QNM quantization scheme from \cite{franke2019quantization} to include cases where the spatial separation of the cavities is large enough for the inter-cavity coupling to be described with the propagating regularized fields. For QNM cavities in a homogeneous background medium, we show that the equal-time inter-cavity overlap becomes small for distant cavities. \red{While coupling in a homogeneous medium is generally weak, and many quantum technologies instead rely on structured environments such as waveguides to connect separated cavities, recent experiments on entanglement distribution using satellites have shown that coupling through homogeneous backgrounds has significant applications \cite{yin2017satellite, yin2017satellite1200}. Furthermore, QNMs in a homogeneous background is a well-established problem \cite{kristensen2012generalized, sauvan2013theory, ge2014quasinormal, franke2019quantization}. Hence, by using a homogeneous background, we focus on the separation of cavities and the definition of multi-cavity operators. A generalization of the quantization scheme to more complex backgrounds is subject to future work.}
We quantify the separation of two cavities \(i\) and \(j\) by defining the cavity separation parameter \(P_{ij}\) that distinguishes cases where the cavities have to be quantized as a joint system and where they can be quantized independently. We test the theory by calculating the overlap for two different systems, first for two coupled 1D dielectric slabs and then for a dimer consisting of two metal nanorods. In both cases, the cavity separation parameter establishes a simple upper bound on the magnitude of the overlap. 
For a sufficiently large separation, i.e., \(P_{ij}\gg 0\), we define QNM operators for the individual cavities. To connect the multi-cavity QNM theory to phenomenological approaches, we reformulate the electromagnetic field Hamiltonian to a system-bath form where separate cavities couple to a shared photonic environment\red{, and which can be used as the starting point for quantum dynamics calculations based on the QNM framework.}

\section{Introduction to optical quasinormal modes} \label{sec:qnmintro}
\begin{figure*}
    \centering
    \includegraphics[width = 0.99\linewidth]{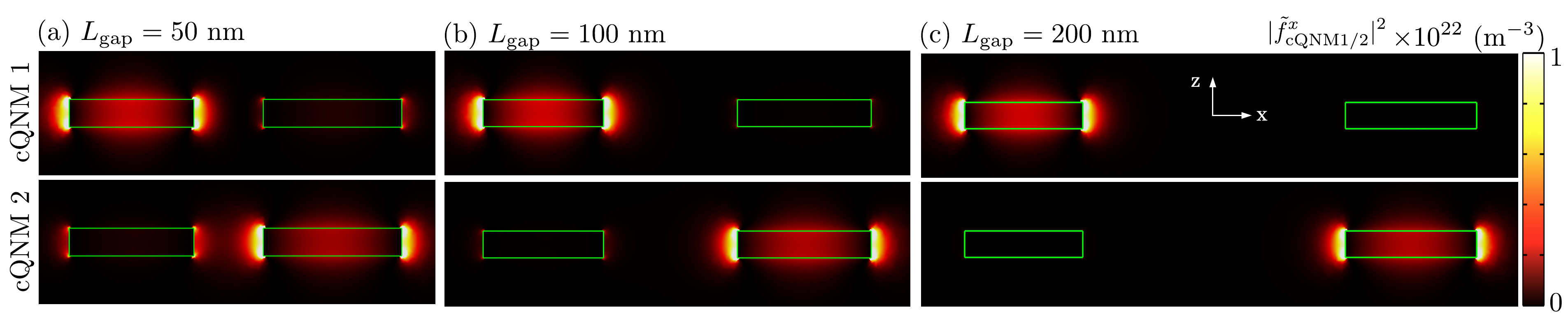}
    \caption{Example of the two dominant QNMs (cQNM 1 and cQNM 2 in the frequency regime of interest) of a dimer consisting of two cylindrical metal nanorods of length \(L_1 = 90\, {\rm nm}\), \(L_2 = 100\,{\rm nm}\) and of diameter \(D_1 = D_2 =20 \, {\rm nm}\) (see Appendix~\ref{appsec:simulationdetail} for more details). The modes are delocalized across the entire dimer for small separations \(L_{\rm gap}\). However, for larger separations, they become increasingly localized at each rod and revert to the mode profile of the isolated rods. }
    \label{fig:dimer}
\end{figure*}

We consider a model of multiple spatially separated optical cavities or plasmonic nanoparticles embedded in a homogeneous background medium. The entire space is divided into the cavity volumes and the outside volume \(V = V_{\rm out}\cup (\cup_i V_i)\), where \(\epsilon(\mathbf{r},\omega)|_{r\in V_{\rm out}} = \epsilon_B\) is the background permittivity. The QNMs of the separate cavities, \(\qnm{f}{i_{\mu}}(\mathbf{r})\), are solutions of the source-free Helmholtz equation \cite{leung1994completeness, muljarov2011brillouin, lalanne2018light, kristensen2020modeling}
\begin{align} \label{eq:helmquasi}
    \nabla\times\nabla\times\Tilde{\mathbf{f}}_{i_{\mu}}(\mathbf{r}) -\frac{\Tilde{\omega}^2_{i_{\mu}}}{c^2}\epsilon(\mathbf{r},\Tilde{\omega}_{i_{\mu}})\Tilde{\mathbf{f}}_{i_{\mu}}(\mathbf{r}) = 0
\end{align}
under open boundary conditions such as the Silver-Müller radiation condition \cite{muller1948grundzuge, silver1984microwave, gumerov2005fast, kristensen2020modeling}
\begin{align} \label{eq:qnmsilvmüll}
    \frac{\mathbf{r}}{r}\times\nabla\times\Tilde{\mathbf{f}}_{i_{\mu}}(\mathbf{r}) \to -in_B\frac{\Tilde{\omega}_{i_{\mu}}}{c}\Tilde{\mathbf{f}}_{i_{\mu}}(\mathbf{r}),\quad r\to \infty
\end{align}
to include the emission of energy into the surrounding medium. Here, \(n_B=\sqrt{\epsilon_B}\) is the refractive index of the background, and \(\epsilon(\mathbf{r},\Tilde{\omega}_{i_{\mu}})\) is the analytic continuation of the electric permittivity (or complex dielectric constant). The mode index \(i_{\mu}\) denotes the \(\mu\)-th mode of the \(i\)-th cavity. Figure~\ref{fig:dimer} shows the field distribution of the dominant QNMs of a dimer of coupled metallic nanorods. For larger distances \(L_{\rm gap}\) between the rods, the coupled QNMs recover the eigenfrequencies and mode profile of the dominant QNMs of the individual rods (cf.~Appendix~\ref{appsec:simulationdetail}). Therefore, in the following, \(\qnm{f}{i_{\mu}}(\mathbf{r})\) is a \textit{single-cavity mode}, i.e., the presence of other cavities is neglected in the Helmholtz equation \red{(setting \(\epsilon(\mathbf{r},\omega) = \epsilon_{\rm B}\) at the other cavities)} or at most treated as a perturbation to the single cavity \cite{franke2023impact}. The present theory is suitable for setups where an individual treatment of the systems is desirable, which is the case for many photon-based quantum information technologies. For strongly coupled systems [e.g., the metallic dimer with small \(L_{\rm gap}\) in Fig.~\ref{fig:dimer}(a)], the existing schemes of hybridized modes and joint operators are the right choice \cite{franke2019quantization, el2020quasinormal, ren2021quasinormal, ren2022connecting, franke2022quantized}. 

Due to the open boundary conditions and for passive media (i.e., no gain), the QNM eigenfrequencies \(\Tilde{\omega}_{i_{\mu}} = \omega_{i_{\mu}} - i\gamma_{i_{\mu}}\) are complex with negative imaginary part \(\gamma_{i_{\mu}}>0\), so that the modes decay with time as is expected for the field inside an open cavity. We assume that the modes \(\qnm{f}{i_{\mu}}(\mathbf{r})\) form a complete set inside the \(i\)-th cavity, permitting an expansion of the Green's function inside the cavity volume \(V_i\) in terms of the QNMs \cite{leung1996completeness, ge2014quasinormal},
\begin{align}\label{eq:qnmgreen}
    \mathbf{G}_{ff}(\mathbf{r},\mathbf{r}',\omega)\Big|_{r,r'\in V_i} = \sum_{\mu}A_{i_{\mu}}(\omega)\qnm{f}{i_{\mu}}(\mathbf{r})\qnm{f}{i_{\mu}}(\mathbf{r}'),
\end{align}
where \(A_{i_{\mu}}(\omega) = \omega/(2(\Tilde{\omega}_{i_{\mu}}-\omega))\). Practical calculations are limited to a frequency interval \(\Delta\omega\) of interest (e.g., the optical regime), where an expansion using a few dominant modes is assumed to give a good approximation of the Green's function. 

As a consequence of the outgoing wave boundary condition [cf. Eq.~\eqref{eq:qnmsilvmüll}] and the complex eigenfrequencies, the QNMs diverge in the far field and hence, are not an efficient basis for positions far away from the resonator. Instead, outside \(V_i\), the QNMs are replaced with regularized fields \cite{ge2014quasinormal, ren2020near} \(\qnm{F}{i_{\mu}}(\mathbf{r},\omega) = \int_{V_i}\mathrm{d}^3r'\, \mathbf{G}_B(\mathbf{r},\mathbf{r}',\omega)\Delta\epsilon(\mathbf{r}',\omega)\qnm{f}{i_{\mu}}(\mathbf{r}')\) (for details, see Appendix \ref{appsec:greenfunc}), where \(\mathbf{G}_B\) is the Green's function of the homogeneous background medium and \(\Delta\epsilon(\mathbf{r},\omega) = \epsilon(\mathbf{r},\omega)-\epsilon_B\). The frequency dependence accounts for the propagation of the fields, and they are non-divergent.

\section{Quasinormal mode quantization}\label{sec:qnmquanti}
In dissipative media, absorption creates a noise current, which, in line with the fluctuation-dissipation theorem, acts as the source of the electric field in the medium\red{. The electric field operator therefore reads} \cite{dung1998three, suttorp2004field, philbin2010canonical, scheel1998qed, matloob1999electromagnetic, PhysRevA.53.1818},
\begin{align} \label{eq:welschquanti}
    \opvec{E}(\mathbf{r},\omega) = \frac{i}{\omega\epsilon_0}\int\mathrm{d}^3r'\mathbf{G}(\mathbf{r},\mathbf{r}',\omega)\cdot\opvec{j}_N(\mathbf{r}',\omega),
\end{align}
where \(\opvec{j}_N(\mathbf{r},\omega)=\omega\sqrt{(\hbar\epsilon_0/\pi)\epsilon_I(\mathbf{r},\omega)}\,\opvec{b}(\mathbf{r},\omega)\) is the noise-current density and \(\opvec{b}(\mathbf{r},\omega)\) are bosonic operators with continuous spatial and frequency indices\red{, i.e.,} 
\begin{align*}
    \red{\left[\hat{b}_i(\mathbf{r},\omega),\hat{b}_j^{\dagger}(\mathbf{r}',\omega')\right] = \delta_{ij}\delta(\mathbf{r}-\mathbf{r}')\delta(\omega-\omega').}
\end{align*} 
Furthermore, \(\epsilon_I(\mathbf{r},\omega)\) is the imaginary part of the permittivity. The Green's function solves the classical Helmholtz equation
\begin{align} \label{eq:helmgreen}
    \left(\nabla\times\nabla\times-\frac{\omega^2}{c^2}\epsilon(\mathbf{r},\omega)\right) \mathbf{G}(\mathbf{r},\mathbf{r}',\omega) = \frac{\omega^2}{c^2}\mathbb{1}\delta(\mathbf{r}-\mathbf{r}'),
\end{align}
where \(\mathbb{1}\) is the identity matrix, and the background Green's function \(\mathbf{G}_B\) (defined earlier) is obtained by solving Eq.~\eqref{eq:helmgreen} for \(\epsilon(\mathbf{r},\omega) = \epsilon_B\).
The frequency-independent electric field operator is obtained by integrating over the frequencies \(\opvec{E}(\mathbf{r})=\int_0^{\infty}\mathrm{d}\omega \opvec{E}(\mathbf{r},\omega)+\mathrm{H.a.}\) 

\begin{figure}
    \centering
    \includegraphics[width=8cm]{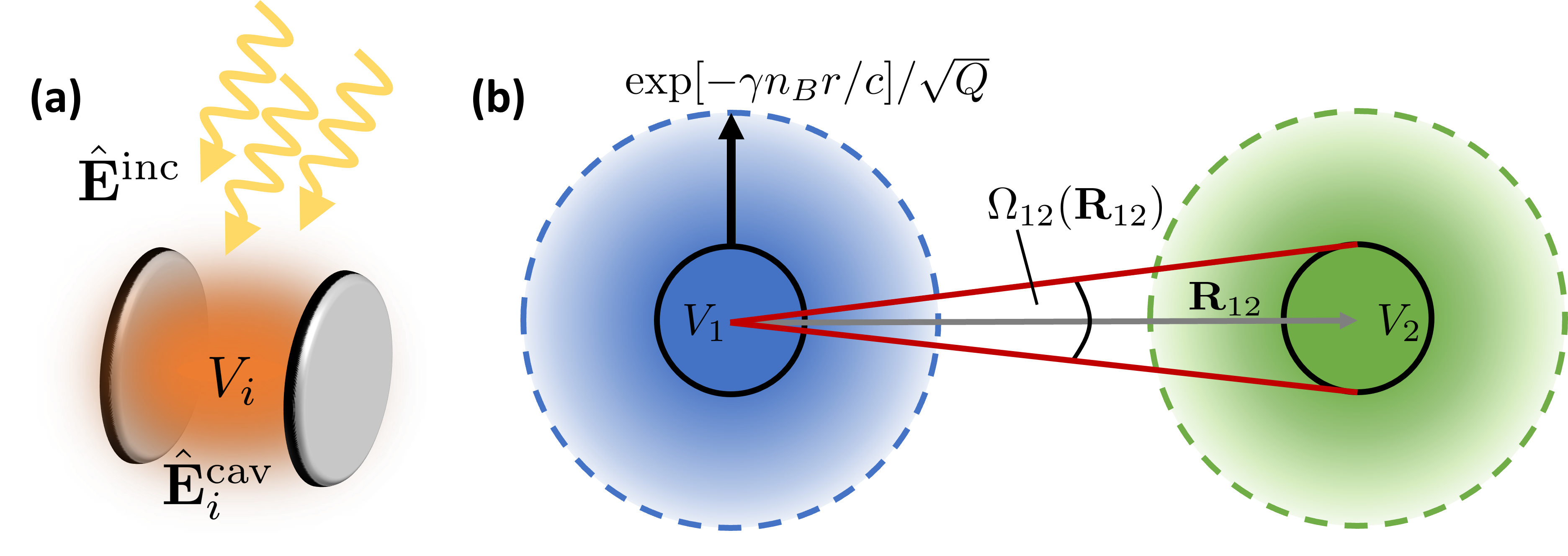}
    \caption{(a) The two contributions to the electric field inside a cavity \(V_i\) are the single cavity field \(\hat{\mathbf{E}}_i^{\mathrm{cav}}\) due to sources inside the cavity, and incoming fields  \(\hat{\mathbf{E}}^{\mathrm{inc}}\) from outside sources (e.g., other cavities). For well-separated cavities, the incoming fields are negligible unless there is a large enough time delay to account for photon propagation. (b) While the QNMs penetrate their cavity boundary, the strength of the cavity field decreases exponentially with the distance from the cavity. The field strength outside relative to the field inside the resonator is weaker by a factor of \(1/\sqrt{Q}\). Two cavities \(1\) and \(2\) can only couple to each other if energy is emitted into the solid angle \(\Omega_{12}\) that connects the cavities, and which, in three dimensions, depends on the distance \(|\mathbf{R}_{12}|\) between the cavities. All of these factors are represented by the cavity separation parameter \(P_{ij}\) [cf.~Eq.~\eqref{eq:cavsepparam}].}
    \label{fig:cav_combined}
\end{figure}

In the multi-cavity model, the sources lie inside the cavity volumes \(V_i\), so that the electric field inside one of the cavities separates into the (single-) cavity field, where the presence of other cavities is not included, and contributions due to sources in the other cavities [cf.~Appendix~\ref{appsec:incfields}, Fig.~\ref{fig:cav_combined}(a)]:
\begin{align}
    \opvec{E}(\mathbf{r})\big|_{r\in V_i} =  \opvec{E}^{\mathrm{cav}}(\mathbf{r})\big|_{r\in V_i}+\opvec{E}^{\mathrm{inc}}(\mathbf{r})\big|_{r\in V_i}.
\label{eq:Edecomp}
\end{align}
The definition of the incoming fields \(\opvec{E}^{\mathrm{inc}}(\mathbf{r})\) is given by
\begin{align}\label{eq:Einc}
    &\opvec{E}^{\mathrm{inc}}(\mathbf{r})\big|_{r\in V_i} = i\sqrt{\frac{\hbar}{\pi\epsilon_0}}\sum_{k\neq i}\int_0^{\infty}\mathrm{d}\omega\int_{V_k}\mathrm{d}^3r'\sqrt{\epsilon_I(\mathbf{r}',\omega)} \nonumber\\
    &\qquad\qquad\qquad\qquad\times\mathbf{G}(\mathbf{r},\mathbf{r}',\omega)\cdot \opvec{b}(\mathbf{r}',\omega) +\mathrm{H.a.}, 
\end{align}
which contain contributions due to incoming radiation from other cavities and non-radiative coupling between the cavities due to absorption in the resonator medium. \red{Equation~\eqref{eq:Einc} contains the full time delay in the dynamics of the noise operators \(\hat{\mathbf{b}}\). The inclusion of the retarded Green's function ensures that the influence of the incoming fields is negligible unless the time delay matches the retardation (see Appendix \ref{appsec:incfields} for a detailed discussion of the incoming fields). For cases where the retardation is small and the influence of the incoming fields to a cavity is large even without time delay, a description in terms of the single cavity modes is generally not useful, and instead, hybrid modes of the joint system have to be used \cite{franke2019quantization, ren2021quasinormal, ren2022connecting}.}

The single cavity field contribution to Eq.~\eqref{eq:Edecomp} reads
\begin{align} \label{eq:Ecav}
    &\opvec{E}^{\mathrm{cav}}(\mathbf{r})\big|_{r\in V_i}\nonumber\\
    &=i\sqrt{\frac{\hbar}{\pi\epsilon_0}}\int_0^{\infty}\mathrm{d}\omega\Big[ \int_{V_i}\mathrm{d}^3r'\sqrt{\epsilon_I(\mathbf{r}',\omega)}\mathbf{G}(\mathbf{r},\mathbf{r}',\omega)\cdot \opvec{b}(\mathbf{r}',\omega) \nonumber\\
    &\qquad+\int_{\overline{V_i}}\mathrm{d}^3r'\sqrt{\epsilon_{B,I}}\mathbf{G}(\mathbf{r},\mathbf{r}',\omega)\cdot \opvec{b}(\mathbf{r}',\omega)\Big] +\mathrm{H.a.},
\end{align}
where \(\overline{V}_i\) is the complement of \(V_i\). Note, that we set \(\epsilon(\mathbf{r},\omega) = \epsilon_B\) everywhere except inside cavity \(i\). This way, \(\opvec{E}^{\mathrm{cav}}(\mathbf{r})\) only contains sources from within that cavity, yielding a separate set of operators for each cavity (a similar separation was performed in \cite{franke2022quantized} for coupled loss-gain resonators). While the background permittivity \(\epsilon_B\) is assumed to be real, it is the limit of a complex permittivity series \red{\(\lim_{\alpha\to 0} (\epsilon_B+\alpha\chi) = \epsilon_B\) for a Kramers-Kronig susceptibility \(\chi\)}. In general, the limit cannot be exchanged with the integration over an infinite volume and needs to be taken last, preserving the fluctuation-dissipation theorem (see \cite{franke2020fluctuation} for details). Instead of writing the limit explicitly, we keep the \red{limit \(\alpha\to 0\)} implicit except for when it is more instructive to write it out. 

By inserting the Green's function expansion using the QNMs of the \(i\)-th cavity [Eq.~\eqref{eq:qnmgreen}] into \(\hat{\mathbf{E}}^{\mathrm{cav}}(\mathbf{r})\), we obtain an expression for the cavity field operator inside the \(i\)-th cavity represented through the QNMs of that cavity,
\begin{align} \label{eq:qnmefield}
    \opvec{E}_i^{\mathrm{cav}}(\mathbf{r}) = i\sqrt{\frac{\hbar}{2\epsilon_0}}\sum_{\mu}\sqrt{\omega_{i_{\mu}}}\qnm{f}{i_{\mu}}(\mathbf{r})\tilde{\alpha}_{i_{\mu}} + \mathrm{H.a.},
\end{align}
where the QNM operators are defined via
\begin{align}
    &\tilde{\alpha}_{i_{\mu}} = \sqrt{\frac{2}{\pi\omega_{i_{\mu}}}}\int_0^{\infty}\mathrm{d}\omega\, A_{i_{\mu}}(\omega)\int\mathrm{d}^3r \, \opvec{b}(\mathbf{r},\omega)\nonumber\\
    &\quad\times\left[\chi_{V_i}(\mathbf{r})\sqrt{\epsilon_I(\mathbf{r},\omega)}\,\qnm{f}{i_{\mu}}(\mathbf{r})+\chi_{\overline{V_i}}(\mathbf{r})\sqrt{\epsilon_{B,I}}\,\qnm{F}{i_{\mu}}(\mathbf{r},\omega)\right],
\end{align}
and \(\chi_V(\mathbf{r})\) is \(1\) if \(r\in V\) and \(0\) if \(r\notin V\). In the single-cavity case, this definition is identical to the quantization from \cite{franke2019quantization}, since outside the cavities \(\epsilon(\mathbf{r},\omega) = \epsilon_B\).
The operators \(\tilde{\alpha}_{i_{\mu}}\) are non-bosonic,
\begin{align} \label{eq:alphacommutator}
     \left[ \tilde{\alpha}_{i_{\mu}}, \tilde{\alpha}^{\dagger}_{j_{\eta}}\right] = S_{i_{\mu}j_{\eta}} \neq \delta_{ij}\delta_{\mu\eta},
\end{align}
where \(S_{i_{\mu}j_{\eta}}\) is the overlap matrix between modes \(i_{\mu}\) and \(j_{\eta}\). The non-bosonic nature is a direct consequence of the open boundary conditions of the QNMs [Eq.~\eqref{eq:qnmsilvmüll}].

We formally divide the overlap matrix \(S_{i_{\mu}j_{\eta}}\) into intra-cavity parts \(\delta_{ij}S^{\rm intra}_{i_{\mu}i_{\eta}}\) and inter-cavity parts \((1-\delta_{ij})S^{\rm inter}_{i_{\mu}j_{\eta}}\).
For the intra-cavity case \(i=j\), the overlap matrix has the same form as for single cavities (cf.~Appendix \ref{appsec:Smatrix}) \cite{franke2019quantization}, 
\begin{align} \label{eq:Sintra}
    S^{\rm intra}_{i_{\mu}i_{\eta}} = S^{\rm nrad}_{i_{\mu}i_{\eta}} + S^{\rm rad}_{i_{\mu}i_{\eta}},
\end{align}
where 
\begin{align}  \label{eq:Snraddef}
    &S^{\rm nrad}_{i_{\mu}i_{\eta}} = \frac{2}{\pi\sqrt{\omega_{i_{\mu}}\omega_{i_{\eta}}}}\int_0^{\infty}\mathrm{d}\omega\, A_{i_{\mu}}(\omega)A^*_{i_{\eta}}(\omega)\nonumber\\
    &\qquad\qquad\times\int_{V_i}\mathrm{d}^3r\epsilon_I(\mathbf{r},\omega)\qnm{f}{i_{\mu}}(\mathbf{r})\cdot\qnm{f}{i_{\eta}}^*(\mathbf{r})
\end{align}
accounts for non-radiative losses via absorption by the cavity medium and 
\begin{align} \label{eq:Sraddef}
    &S^{\rm rad}_{i_{\mu}i_{\eta}} = \frac{2}{\pi\sqrt{\omega_{i_{\mu}}\omega_{i_{\eta}}}}\int_0^{\infty}\mathrm{d}\omega\, \frac{A_{i_{\mu}}(\omega)A^*_{i_{\eta}}(\omega)}{2\omega \epsilon_0}\nonumber\\
    &\times\oint_{\mathcal{S}_i}\mathrm{d}A_s\left[\left(\qnm{H}{i_{\mu}}(\mathbf{s},\omega)\times\opvec{n}_s\right)\cdot\qnm{F}{i_{\eta}}^*(\mathbf{s},\omega)+\mathrm{c.c.}(\mu\leftrightarrow \eta)\right]
\end{align}
is associated with radiative losses through the cavity surface \(\mathcal{S}_i\), with the normal vector \(\opvec{n}_s\) pointing outwards with respect to the cavity volume \(V_i\). 

We have defined the regularized magnetic field QNMs via \(\qnm{H}{i_{\mu}}(\mathbf{r},\omega) = -i\nabla\times\qnm{F}{i_{\mu}}(\mathbf{r},\omega)/(\omega\mu_0)\). We highlight that the expression of the overlap matrix using surface integrals [cf.~Eq.~\eqref{eq:Sraddef}] greatly reduces the numerical cost of calculating the overlap, compared to the full integral over the volume outside the cavities (cf. Appendix \ref{appsec:Smatrix}).
As shown in Appendix \ref{appsec:Smatrix}, the inter-cavity overlap matrix reads
\begin{widetext}
\begin{align} \label{eq:Sinterdef}
    S^{\rm inter}_{i_{\mu}j_{\eta}}\Big|_{i\neq j} = \frac{2}{\pi\sqrt{\omega_{i_{\mu}}\omega_{j_{\eta}}}}\int_0^{\infty}\mathrm{d}\omega\,\frac{A_{i_{\mu}}(\omega)A^*_{j_{\eta}}(\omega)}{2\omega \epsilon_0}&\bigg\{\oint_{\mathcal{S}_i}\mathrm{d}A_s\left[\left(\qnm{H}{i_{\mu}}(\mathbf{s},\omega)\times\opvec{n}_s\right)\cdot\qnm{F}{j_{\eta}}^*(\mathbf{s},\omega)+\mathrm{c.c.}(i_{\mu}\leftrightarrow j_{\eta})\right]\nonumber\\
    &+\oint_{\mathcal{S}_j}\mathrm{d}A_s\left[\left(\qnm{H}{i_{\mu}}(\mathbf{s},\omega)\times\opvec{n}_s\right)\cdot\qnm{F}{j_{\eta}}^*(\mathbf{s},\omega)+\mathrm{c.c.}(i_{\mu}\leftrightarrow j_{\eta})\right]\bigg\}.
\end{align}
\end{widetext}
The definitions of the intra- and inter-cavity overlap matrices are noticeably different. Since absorption from other cavities was not included in the quantization of the individual cavity but moved to a separate contribution, the intra-cavity overlap is identical to the single-cavity terms from \cite{franke2019quantization, franke2020fluctuation}. The inter-cavity overlap, meanwhile, seems to resemble the radiative part \(S^{\rm rad}\) from the single-cavity case but now involves the surfaces of two separate cavities. It thus includes the net exchange of energy through the two cavity surfaces. On the other hand, even for far-away cavities, there is also a non-radiative contribution due to absorption by the resonator material. An inter-cavity term completely analogous to \(S^{\rm nrad}\) is not present here, though, since the influence of absorptive losses in other cavities on the (single-)cavity field was moved into the incoming fields \(\hat{\mathbf{E}}^{\rm inc}(\mathbf{r})\) from Eq.~\eqref{eq:Einc}.

Note, however, that \(S_{i_{\mu}j_{\eta}}\)  is an equal-time quantity and does not include propagation between separated cavities. We expect the inter-cavity overlap to vanish for perfectly localized modes unless the time delay is larger or equal to the propagation time between the cavities. 
\red{But the open boundary conditions [cf. Eq.~\eqref{eq:qnmsilvmüll}] lead to a finite linewidth of the QNMs. Therefore, as a consequence of energy-time uncertainty,} the inter-cavity overlap remains finite even without time delay. However, the separation still results in a small, often negligible, overlap between far-away cavities. 

To derive a quantitative measure of this overlap, we assume that the QNM wavelengths are small compared to their spatial separation so that the regularized fields \(\qnm{F}{j_{\eta}}^*(\mathbf{r},\omega)\) [or \(\qnm{H}{j_{\eta}}^*(\mathbf{r},\omega)\)] in the \(\mathcal{S}_i\)-integral of  Eq.~\eqref{eq:Sinterdef} are outgoing waves of the form
\begin{align}\label{eq:Fdecomp}
    \qnm{F}{j_{\eta}}^*(\mathbf{r},\omega)\big|_{r\in V_i} = \qnm{F}{j_{\eta}}'^*(\mathbf{r},\omega)\mathrm{e}^{-i\omega \tau_{ij}},
\end{align}
where \(\tau_{ij} = n_BR_{ij}/c\) is the travel time of photons through the bath over the distance \(R_{ij}\) between the cavity centers [cf. Fig.~\ref{fig:cav_combined}(b)], which is assumed to be large against the size of the cavities. 
We perform the same approximation on the field \(\qnm{F}{i_{\mu}}(\mathbf{r},\omega)\) [\(\qnm{H}{i_{\mu}}(\mathbf{r},\omega)\)] in the \(\mathcal{S}_j\)-integral. Then, we evaluate the frequency integral using the residue theorem on the poles contained in \(A_{i_{\mu}}(\omega)\) and \(A_{j_{\eta}}^*(\omega)\). Since the envelope functions \(\qnm{F}{j_{\eta}}'^*(\mathbf{r},\omega)\) vary slowly with the frequency, their contribution to the frequency integral is negligible. 
It thus follows that the overlap matrix element \(S^{\rm inter}_{i_{\mu}j_{\eta}}\) decreases as \(\mathcal{O}(\mathrm{e}^{-\gamma_{i_{\mu}j_{\eta}}^{\min}\tau_{ij}})\) with the distance \(R_{ij}\) between the cavities (cf.~Appendix \ref{appsec:cavsepparam}), where, \(\gamma_{i_{\mu}j_{\eta}}^{\min}=\min(\gamma_{i_{\mu}},\gamma_{j_{\eta}})\).

On the other hand, high-\(Q\) modes, where \(Q_{i_{\mu}} = \omega_{i_{\mu}}/(2\gamma_{i_{\mu}})\) is the quality factor of the QNM, lose energy at a lower rate than low-quality modes, leading to a weaker coupling to other cavities, and therefore to a decrease in the inter-cavity overlap (relative to the intra-cavity contributions).
For high-Q modes, the amplitude of the regularized fields \(\qnm{F}{i_{\mu}}(\mathbf{r},\omega)\) outside the cavity decreases on the order \(\mathcal{O}(1/\sqrt{Q_{i_{\mu}}})\) (see Appendix \ref{appsec:cavsepparam} for details).  Hence, \(S^{\rm inter}_{i_{\mu}j_{\eta}}\) decreases in this limit like \((Q_{i_{\mu}}Q_{j_{\eta}})^{-1/2}\).

\begin{figure}
    \centering
    \includegraphics[width = 0.95\columnwidth]{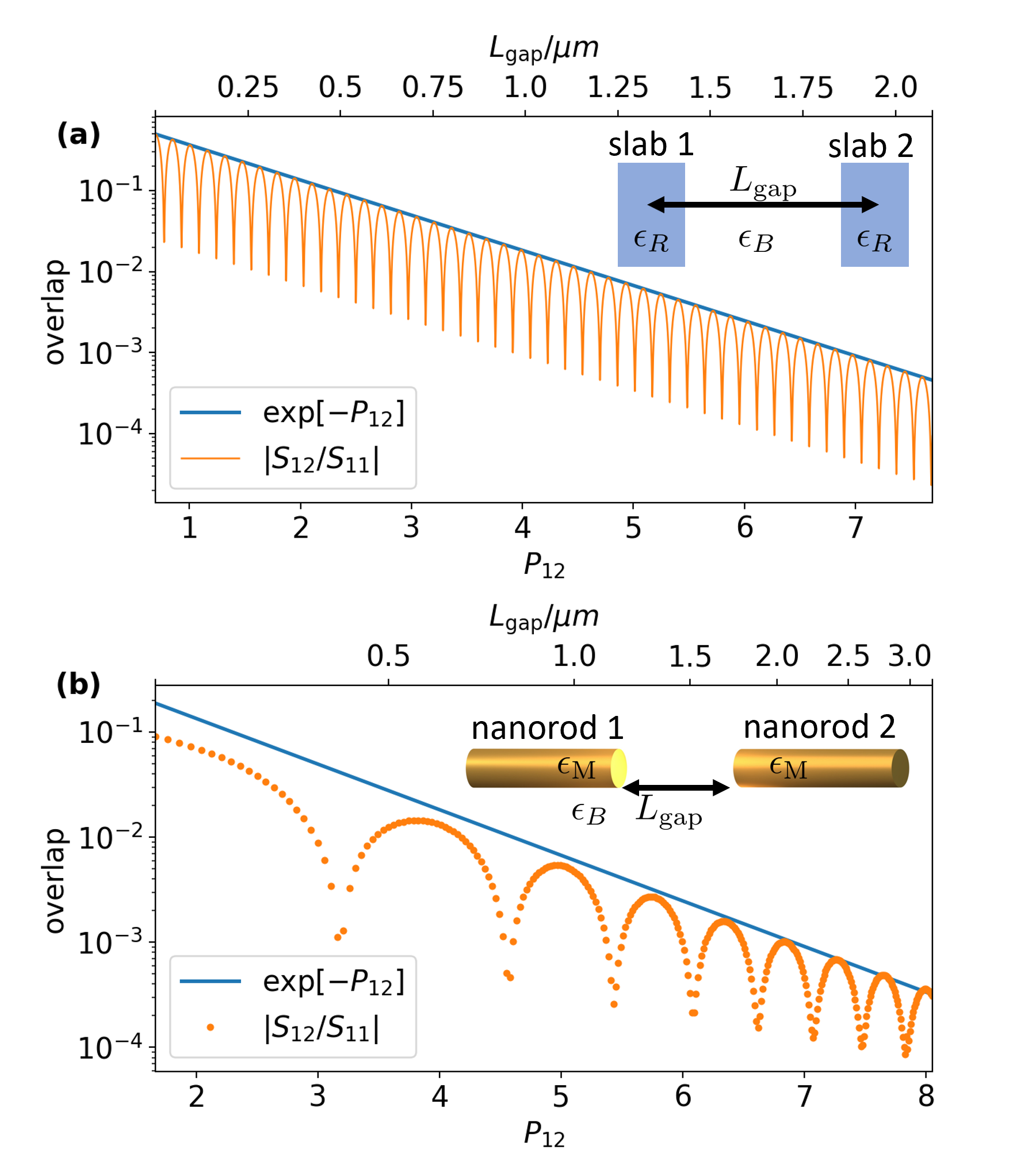}
    \caption{ Inter-cavity overlap for two different systems as a function of the cavity separation parameter \(P_{12}\) and the inter-cavity distance \(L_{\rm gap}\). The exponential scaling with \(P_{12}\) is highlighted. (a) Example of two coupled 1D dielectric slabs with \red{refractive index \(n_{\rm R} = \sqrt{\epsilon_{\rm R}} = 12.7\) in a vacuum background}. The slabs are identical and are assumed to be dominated by a single QNM each, with \(\Tilde{\omega}_1 = \Tilde{\omega}_2\), \(Q = \omega_1/(2\gamma_1) = 10\), and wavelength \(\lambda = 1150\,{\rm nm}\). The exponential scaling with \(P_{12}\) gives a perfect upper bound on the magnitude of the overlap. Oscillations in the overlap are a result of phase differences for varying distances. (b) Overlap for two metal nanorods \red{in vacuum} with \red{Drude} permittivity \(\epsilon_{\rm M}\) and length \(L_1 = 90\, {\rm nm}\) and \(L_2 = 100\, {\rm nm}\) with one dominant QNM each. The QNM wavelengths are \(\lambda_1 = 700\, {\rm nm}\) and \(\lambda_2 = 750\, {\rm nm}\), respectively \red{(cf.~Appendix~\ref{appsec:simulationdetail})}.}
    \label{fig:overlapquant}
\end{figure}

Finally, the directionality of the cavity emission also plays a major role in the overlap. For example, for the two one-dimensional dielectric slabs in Fig.~\ref{fig:overlapquant}(a), only half of the emitted radiation from slab 1 reaches slab 2, while the other half is emitted away from the second slab. 
Consequently,  \(S^{\rm inter}_{i_{\mu}j_{\eta}}\) scales as \(D_{i_{\mu}j_{\eta}}^{\max} = \max[D_{i_{\mu}}(\Omega_{ij}), D_{j_{\eta}}(\Omega_{ji})]\), where \(D_{i_{\mu}}(\Omega_{ij})\) is the fraction of the total radiated power by the field of mode \(i_{\mu}\) that is radiated into the solid angle \(\Omega_{ij}\) towards the other cavity [cf.~Fig.~\ref{fig:cav_combined}(b)], i.e., the directivity \cite{balanis2016antenna} of the emission in this direction (cf. Appendix~\ref{appsec:cavsepparam}). If the emission occurs entirely towards the other cavity, \(D_{i_{\mu}}(\Omega_{ij})=1\). If the emission occurs away from the other cavity, \(D_{i_{\mu}}(\Omega_{ij})=0\). Note that, in three dimensions, \(\Omega_{ij}(\mathbf{R}_{ij})\) is a function of the distance between the cavities and decreases with increasing separation.

We combine these effects into the exponential scaling \(S^{\rm inter}_{i_{\mu}j_{\eta}}\sim \mathrm{e}^{-P_{i_{\mu}j_{\eta}}}\) with the separation parameter 
\begin{align} \label{eq:cavsepparam}
    P_{i_{\mu}j_{\eta}} = \gamma_{i_{\mu}j_{\eta}}^{\min}n_BR_{ij}/c + \frac{1}{2}\mathrm{ln}\left(Q_{i_{\mu}}Q_{j_{\eta}}\right)-\mathrm{ln}(D_{i_{\mu}j_{\eta}}^{\max})\red{.}
\end{align}
The parameter \(P_{i_{\mu}j_{\eta}}\) allows the distinction between cases with significant or negligible overlap between modes of different cavities. \red{Unlike other methods of testing separation, such as the stability of the eigenvalues of the system (cf.~Appendix~\ref{appsec:simulationdetail}) or calculating the full overlap integrals, \(P_{i_{\mu}j_{\eta}}\) depends only on few basic properties of the QNM cavities that are readily available, e.g. for experimental groups, and gives a simple, accessible analytic estimate of the overlap.} The first term in Eq.~\eqref{eq:cavsepparam} characterizes the spatial separation, while the second term accounts for the vanishing overlap of high-quality cavity modes, and the third term contains the shape of the emission and relative orientation of the cavities. A separate quantization of two cavities \(i\) and \(j\) is possible if \(P_{ij} \equiv \min_{\mu\eta}(P_{i_{\mu}j_{\eta}})\gg 0\) for all relevant \(\mu,\eta\) within the frequency regime of interest (cf.~Appendix \ref{appsec:cavsepparam}). Therefore, we refer to \(P_{ij}\) as the \textit{cavity separation parameter}. 

An example of the classification is given in Fig.~\ref{fig:overlapquant}(a) for two identical 1D dielectric slabs (\(1\) and \(2\)) serving as QNM cavities (for a derivation of the overlap matrix, see Ref.~\cite{fuchs2023hierarchical}). For this example, with one dominant QNM each, the inter-cavity overlap can be calculated analytically. The exponential scaling with the cavity separation parameter establishes a perfect upper limit for the overlap for this simple case.

A more practical example is that of two coupled metal cylindrical nanorods in free space (cf.~inset of Fig.~\ref{fig:overlapquant} (b)). As shown in Appendix~\ref{appsec:simulationdetail}, these two nanorods have the same diameter ($D_1=D_2=20~$nm) and different lengths ($L_1=90~$nm, $L_2=100~$nm). The surface-to-surface gap distance $L_{\rm gap}$ ranges from $200~$nm to $3200~$nm. The dielectric function of nanorod 1 is governed by the Drude model (see Eq.~\eqref{eq:Drude}). As for nanorod 2, a similar Drude model is used but with a reduced damping rate. A dipole technique~\cite{bai_efficient_2013-1} is used to get the bare QNMs for isolated metal nanorods. 
The overlap is calculated using a pole approximation method by applying the residue theorem to Eq.~\eqref{eq:Sinterdef}, and drastically reducing the numerical cost by eliminating the frequency integration (cf.~Appendix~\ref{appsec:poleapprox}). Combined with the fact that only the QNMs of the isolated rods are required for the new method, the overlap calculation is very efficient (see Appendix~\ref{appsec:simulationdetail} for more details). Figure~\ref{fig:overlapquant}(b) shows the overlap as a function of the cavity separation parameter. The directivity of the rods is approximated for a simple dipole emission so that the exponential scaling with \(P_{12}\) matches the overlap better for larger separations, where higher orders are negligible. \red{The exponential scaling deviates increasingly from the numerical results for very small distances. This is likely due to a breakdown of the assumptions made in the derivation of \(P_{12}\), mainly assuming a sufficiently large separation of the cavities. Thus, the difference between the numerical results and scaling with \(P_{12}\) at very small separations is an indication that the theory breaks down in this regime, and hybridized modes or coupled QNM theory have to be used instead \cite{franke2019quantization, ren2022connecting}.} As in the 1D case, the overlap shows oscillations due to phase differences between the QNMs for different distances between the rods. Still, the cavity separation Parameter \(P_{12}\) establishes an upper bound on the overlap using only a few readily accessible QNM parameters. \red{Thus, the relative inter-cavity overlap can be connected to a specific value of the separation parameter. For example, a value of \(P_{12}\geq 3.0\) ensures an overlap of less than  \(5\%\), while \(P_{12}\geq 4.7\) gives a relative overlap of less than \(1\%\).}

\red{In Fig.~\ref{fig:Purcell}, we also show the Purcell factor (see Appendix~\ref{appsec:Purcell} for details on the calculation) \cite{purcell1946resonance, purcell1946spontaneous}
\begin{align}
    F_{\rm P} = \frac{\Gamma}{\Gamma_0}
\end{align}
for a two-level system near rod 1 and for two different distances between the rods, \(L_{\rm gap} = 100\,{\rm nm}\) (\(P_{12} \approx 0.25\)) and \(L_{\rm gap} = 500\,{\rm nm}\) (\(P_{12}\approx 3.58\)). Here, \(\Gamma\) is the full cavity-enhanced spontaneous emission rate, while \(\Gamma_0\) is the spontaneous emission in the homogeneous background medium, which are given in Eq.~\eqref{appeq:spemcav} and below Eq.~\eqref{eq:Gamma_QNM}, respectively.
It can be seen that, for a distance of \(L_{\rm gap} = 500\,{\rm nm}\), the quantum Purcell factor for two separated rods matches very well with the numerical calculations of the classical Purcell Factor. Furthermore, the Purcell factor with and without the inclusion of rod 2 match very well since well-separated rods are assumed in the derivation (cf.~Appendix~\ref{appsec:Purcell}). This assumption does not hold for \(L_{\rm gap} = 100\,{\rm nm}\) so that the results show some deviations from the numerical calculations, especially near the resonance frequency of the second rod. This matches the results from Fig.~\ref{fig:overlapquant}, where, for a distance of \(L_{\rm gap} = 500\,{\rm nm}\), the overlap between the QNMs of the two rods is small, and the presence of rod 2 only has a small influence on the emission of a dipole near rod 1. For \(L_{\rm gap} = 100\,{\rm nm}\), however, the overlap of the QNMs is significant, and hybridized modes or coupled QNM theory \cite{franke2019quantization, ren2022connecting} have to be used instead to obtain the correct Purcell factors.}

\begin{figure}
    \centering
    \includegraphics[width=\linewidth]{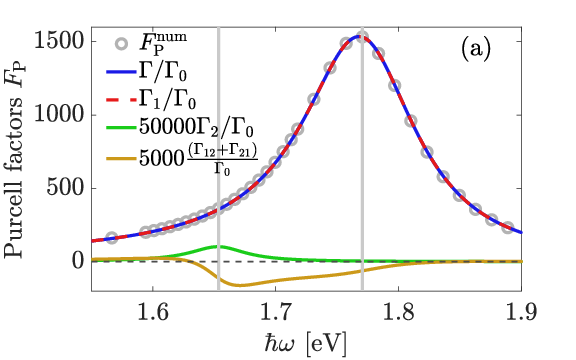}
    \quad
    \includegraphics[width=\linewidth]{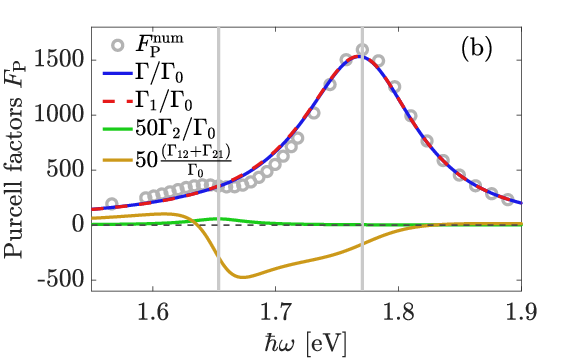}
    \caption{\red{Purcell 
    factors for a system of two gold-like nanorods (cf.~Appendix~\ref{appsec:simulationdetail}) and a dipole \(10\,\)nm away from rod 1 for a distance between the rods of (a) \(L_{\rm gap} = 500\, {\rm nm}\) and (b) \(L_{\rm gap} = 100\, {\rm nm}\). The grey circles mark full numerical calculations of the classical Purcell factor as a reference. The solid blue line shows the quantum Purcell factor as discussed in Appendix.~\ref{appsec:Purcell}, while the dashed red line shows just the Purcell factor due to the coupling of the dipole to the dominant QNM of rod 1. The solid green curve shows the coupling to the dominant QNM of rod 2, while the dashed orange curve shows the influence of coupling between the modes. The vertical lines mark the positions of the resonance of the dominant QNMs of each of the rods.}}
    \label{fig:Purcell}
\end{figure}

\red{These examples show that}, for sufficiently separated cavities, \red{the overlap between modes of different cavities is negligible, i.e.,} the overlap matrix is a diagonal block matrix \(S_{i_{\mu}j_{\eta}} = \delta_{ij} S_{i_{\mu}i_{\eta}}\). Using that \(S_{i_{\mu}i_{\eta}}\) is a Gramian matrix, a Löwdin transformation \cite{lowdin1950non} is performed within each cavity to obtain bosonic cavity operators,
\begin{align} \label{eq:opsymm}
    \hat{a}_{i_{\mu}} = \sum_{\eta} (S^{-1/2})_{i_{\mu}i_{\eta}}\tilde{\alpha}_{i_{\eta}}, 
\end{align}
for the symmetrized QNMs,
\begin{equation}
\qnm{f}{i_{\mu}}^s(\mathbf{r}) = \sum_{\eta}(S^{1/2})_{i_{\eta}i_{\mu}}\sqrt{\omega_{i_{\eta}}/\omega_{i_{\mu}}}\, \qnm{f}{i_{\eta}}(\mathbf{r}),
\end{equation}
of each individual cavity. Using the bosonic operators, we can construct Fock states for the separate systems \cite{franke2019quantization}. \red{The Fock state of the joint system is a product state of the Fock states of the separated systems.} If the separation between two cavities is small, however, the cavities cannot be treated as independent systems. In this respect, the overlap of open cavity modes is similar to the overlap of atomic orbitals, where, if two atoms get too close to each other, a description in terms of the orbitals of the isolated atoms becomes insufficient, and the system has to be described as a molecule. 

\section{The quasinormal mode Hamiltonian}\label{sec:qnmham}
Phenomenological approaches to problems involving 
spatially separated coupled cavities usually start with a Hamiltonian, where separate systems, described using commuting sets of operators, interact with their surroundings \cite{carmichael1993quantum, imamog1999quantum, cirac1997quantum, pellizzari1997quantum}. \red{We have discussed the construction of commuting operators for separate systems in the previous section.} Towards a thorough definition of the Hamiltonian, the bosonic QNM operators are rewritten as a projection of the bosonic noise operators  \(\hat{\mathbf{b}}(\mathbf{r},\omega)\) onto the QNM subspace \cite{franke2020quantized}
\begin{align*}
    \hat{a}_{i_{\mu}} = \int_0^{\infty}\text{d}\omega\int\text{d}^3r\, \mathbf{L}_{i_{\mu}}(\mathbf{r},\omega)\cdot\opvec{b}(\mathbf{r},\omega),
\end{align*}
where 
\begin{align} \label{eq:Ldef}
    &\mathbf{L}_{i_{\mu}}(\mathbf{r},\omega) = \sum_{\eta}\left(S^{-1/2}\right)_{i_{\mu}i_{\eta}}\sqrt{\frac{2}{\pi\omega_{i_{\eta}}}}A_{i_{\eta}}(\omega)\nonumber\\
    &\qquad\times\left[\chi_{V_i}(\mathbf{r})\sqrt{\epsilon_I(\mathbf{r},\omega)}\,\qnm{f}{i_{\eta}}(\mathbf{r})+\chi_{\overline{V_i}}(\mathbf{r})\sqrt{\epsilon_{B,I}}\,\qnm{F}{i_{\eta}}(\mathbf{r},\omega)\right],
\end{align}
are QNM projector kernels. These are orthogonal by construction, i.e., \(\int\mathrm{d}^3r\int_0^{\infty}\mathrm{d}\omega \mathbf{L}_{i_{\mu}}(\mathbf{r},\omega)\cdot\mathbf{L}^*_{j_{\eta}}(\mathbf{r},\omega) = \delta_{ij}\delta_{\mu\eta}\). Hence, we decompose the noise operators \(\opvec{b}\) into a system and bath part,
\begin{align}\label{eq:bdecomp}
    \opvec{b}(\mathbf{r},\omega) = \sum_i\sum_{\mu}\mathbf{L}_{i_{\mu}}^*(\mathbf{r}, \omega)\hat{a}_{i_{\mu}} + \opvec{c}(\mathbf{r},\omega),
\end{align}  
where QNM and bath operators commute. Thus, the full Hamiltonian of the electromagnetic field  \(H = \hbar \int\mathrm{d}^3r\int_0^{\infty}\mathrm{d}\omega\, \omega \hat{\mathbf{b}}^{\dagger}(\mathbf{r},\omega)\hat{\mathbf{b}}(\mathbf{r},\omega)\) \cite{dung1998three} transforms into a system-bath form \cite{franke2020quantized}:
\begin{align}\label{eq:qnmham}
    &H = H_S + H_B + H_{SB}\nonumber\\
    &= \hbar\sum_{i,\mu\eta}\left(\int_0^{\infty}\text{d}\omega\int\text{d}^3r\, \omega\, \mathbf{L}_{i_{\mu}}(\mathbf{r},\omega)\cdot\mathbf{L}^*_{i_{\eta}}(\mathbf{r},\omega)\right)\hat{a}^{\dagger}_{i_{\mu}}\hat{a}_{i_{\eta}}\nonumber\\
    &\quad+ \hbar\int_0^{\infty}\text{d}\omega\int\text{d}^3r\, \omega\,\opvec{c}^{\dagger}(\mathbf{r},\omega)\cdot\opvec{c}(\mathbf{r},\omega)\nonumber\\
    &\quad + \hbar\sum_{i,\mu}\int_0^{\infty}\text{d}\omega\int\text{d}^3r\, \omega\mathbf{L}_{i_{\mu}}(\mathbf{r},\omega)\cdot\opvec{c}(\mathbf{r},\omega)\,\hat{a}_{i_{\mu}}^{\dagger} + \text{H.a.}
\end{align}
The first term in Eq.~\eqref{eq:qnmham} includes no coupling between well-separated systems 
[cf.~Eq.~\eqref{eq:cavsepparam}]. Thus, there is no direct (meaning instantaneous) coupling between the QNMs of far-away cavities contained in the Hamiltonian. Note, however, that the off-diagonal terms (\(\mu\neq \eta\)) contain scattering between modes of the same cavity and between modes of strongly coupled systems with negligible separation. Thus, the previous QNM theory is still contained in the limit of strongly coupled cavities. Furthermore, the QNMs of all cavities are coupled to the same bath. This will lead to an effective inter-cavity coupling if the time delay is large enough to permit photon propagation between the cavities\red{, so that the causality respecting energy transfer between separated systems is fully included in the quantized theory.}

For cases where off-diagonal coupling between modes is negligible, the system-bath Hamiltonian can be brought into the familiar form by combining the coupling elements and bath operators into Gaussian noise operators related to emission/absorption of cavity photons from mode \(i_{\mu}\) (see Appendix~\ref{appsec:pheno} for the derivation). Transforming these noise operators into bosonic bath operators \(\hat{B}_{i_{\mu}}\) yields the following Hamiltonian for the system-bath coupling
\begin{align}
    H_{SB} = i\hbar \sum_{i,\mu}\sqrt{2\gamma_{i_{\mu}}}\hat{a}^{\dagger}_{i_{\mu}}\hat{B}_{i_{\mu}}+\mathrm{H.a.},
\end{align}
which is the same form as in the model from \cite{carmichael1993quantum}, demonstrating that the common phenomenological models are obtained as a limit of the QNM theory for certain geometries with sufficiently high $Q$ cavities. Note, however, that in some cases, even high-$Q$ cavities can exhibit non-Hermitian overlap~\cite{ren2022connecting}, and a description using quantized QNMs is needed. 

\section{Conclusions}
We have presented an extension of the established QNM quantization scheme to systems of multiple, spatially separated cavities in a homogeneous background medium. We showed that, while the QNMs penetrate their resonator boundary and extend into the surrounding medium, an individual quantization of the cavity modes is possible, even for highly dissipative systems, as long as the inter-cavity mode overlap is small. We quantified this by defining a cavity separation parameter \(P_{ij}\) that accounts for the spatial separation of the cavities and the quality factor of the cavity modes. \red{The parameter \(P_{ij}\) thus connects to the maximum possible value of the overlap for a specific geometry.} For the example of two identical dielectric slabs (\(1\) and \(2\)) dominated by a single QNM each, the inter-cavity overlap of the modes dropped below \red{five percent} of the intra-cavity contributions for \red{\(P_{12}\geq 3.0\)}, and below one percent for \(P_{12} \geq 4.7\). The same behavior has been demonstrated for the example of a coupled system of two three-dimensional metal nanorods (with slightly different permittivities) In this coupling regime, the cavities can be modeled as separate systems that couple to a shared photonic bath, which gives rise to effective cavity-cavity interactions if the time delay is significant enough to allow for photon propagation times.

\begin{acknowledgments}
R.F. and M.R. acknowledge support from the Deutsche Forschungsgemeinschaft (DFG) - Project number 525575745.
S.H. acknowledges funding from Queen's University, Canada, 
the Canadian Foundation for Innovation (CFI), 
the Natural Sciences and Engineering Research Council of Canada (NSERC), CMC Microsystems for the provision of COMSOL Multiphysics, and 
the 
Alexander von Humboldt Foundation through a Humboldt Award.

\end{acknowledgments}

\appendix
\section{Green's function expansion in terms of QNMs for multiple cavities}\label{appsec:greenfunc}
Using the Green's function Helmholtz equation [Eq.~\eqref{eq:helmgreen}], we find the following Dyson equation for the full Green's function \cite{ge2014quasinormal}
\begin{align}\label{appeq:genGreenDyson}
    &\mathbf{G}(\mathbf{r},\mathbf{r}',\omega) = \mathbf{G}_B(\mathbf{r},\mathbf{r}',\omega) \nonumber\\
    &+ \int\mathrm{d}^3r''\mathbf{G}_B(\mathbf{r},\mathbf{r}'',\omega)\Delta\epsilon(\mathbf{r}'',\omega)\mathbf{G}(\mathbf{r}'',\mathbf{r}',\omega),
\end{align}
where \(\Delta\epsilon(\mathbf{r},\omega) = \epsilon(\mathbf{r},\omega)-\epsilon_B\) vanishes for positions inside the homogeneous background medium. Assuming \(r'\in V_i\) and \(r\in V_{\rm out}\), we insert the expansion from Eq.~\eqref{eq:qnmgreen} to obtain 
\begin{align} \label{appeq:GreenDyson}
    &\mathbf{G}(\mathbf{r},\mathbf{r}',\omega)\Big|_{r'\in V_i}\nonumber\\
    &\qquad= \mathbf{G}_B(\mathbf{r},\mathbf{r}',\omega) + \sum_{\mu}A_{i_{\mu}}(\omega)\qnm{F}{i_{\mu}}(\mathbf{r},\omega)\qnm{f}{i_{\mu}}(\mathbf{r}')\nonumber\\
    &\qquad +\sum_{k\neq i}\int_{V_k}\mathrm{d}^3r''\mathbf{G}_B(\mathbf{r},\mathbf{r}'',\omega)\Delta\epsilon(\mathbf{r}'',\omega)\mathbf{G}(\mathbf{r}'',\mathbf{r}',\omega).
\end{align}
The first line on the right-hand side is identical to the single-cavity case in \cite{ge2014quasinormal}, the first part being the unperturbed background Green's function, while in the second part \(\qnm{F}{i_{\mu}}(\mathbf{r},\omega)\) is the scattered field at cavity \(i\).
Unlike the single-cavity case, we obtain a third term here due to scattering at the other cavities. While it is generally possible to include these higher-order scattering terms, doing so comes at additional numerical cost and may be unfeasible for complicated structures. We neglect them here for far-away cavities. For cavities with a small spatial separation, where the inclusion of higher-order terms might be necessary, the single-cavity theory with hybridized modes \cite{franke2020quantized} or QNM coupled mode theory \cite{ren2021quasinormal} provide a feasible non-perturbative approach to strongly coupled resonators. 

\section{Incoming fields}\label{appsec:incfields}
We divide the electric field inside cavity \(i\) [cf. Eq.~\eqref{eq:welschquanti}] by adding a zero
\begin{align} \label{appeq:efieldeffzero}
    &\opvec{E}(\mathbf{r})\Big|_{r\in V_i} = \int_0^{\infty}\mathrm{d}\omega\frac{i}{\omega\epsilon_0}\int\mathrm{d}^3r'\mathbf{G}(\mathbf{r},\mathbf{r}',\omega)\cdot\opvec{j}_N(\mathbf{r}',\omega)+\mathrm{H.a.}\nonumber\\
    &=i\sqrt{\frac{\hbar}{\pi\epsilon_0}}\int_0^{\infty}\mathrm{d}\omega\Bigg[ \int_{V_i}\mathrm{d}^3r'\sqrt{\epsilon_I(\mathbf{r}',\omega)}\mathbf{G}(\mathbf{r},\mathbf{r}',\omega)\cdot \opvec{b}(\mathbf{r}',\omega)\nonumber\\
    &\qquad\qquad\qquad+\int_{\overline{V_i}}\mathrm{d}^3r'\sqrt{\epsilon_{B,I}}\mathbf{G}(\mathbf{r},\mathbf{r}',\omega)\cdot \opvec{b}(\mathbf{r}',\omega) \nonumber\\
    &+\int_{\overline{V_i}}\mathrm{d}^3r'\left(\sqrt{\epsilon_I(\mathbf{r}',\omega)}-\sqrt{\epsilon_{B,I}}\right)\mathbf{G}(\mathbf{r},\mathbf{r}',\omega)\cdot \opvec{b}(\mathbf{r}',\omega) \Bigg]\nonumber\\
    &\qquad\qquad\qquad\qquad\qquad+\mathrm{H.a.},
\end{align}
so the absorption in other cavities is contained only in the last term. This way, the first two terms in the square brackets represent the (single-) cavity field \(\opvec{E}^{\mathrm{cav}}(\mathbf{r})\) [cf. Eq.~\eqref{eq:Ecav}], of an isolated cavity in a homogeneous background medium \cite{franke2019quantization}. By inserting the Green's function expansion in terms of the QNMs of that cavity into \(\opvec{E}^{\mathrm{cav}}(\mathbf{r})\), we obtain the expression of the electric field in terms of QNMs from Eq.~\eqref{eq:qnmefield}.

The last term inside the square brackets in Eq.~\eqref{appeq:efieldeffzero} represents the field contributions due to absorption in other cavities and reads,
\begin{align}\label{appeq:incomingfields}
    &\opvec{E}^{\mathrm{inc}}(\mathbf{r})\Big|_{r\in V_i} \nonumber\\
    & \quad=i\sqrt{\frac{\hbar}{\pi\epsilon_0}}\int_0^{\infty}\mathrm{d}\omega\sum_{k\neq i}\int_{V_k}\mathrm{d}^3r'\left(\sqrt{\epsilon_I(\mathbf{r}',\omega)}-\sqrt{\epsilon_{B,I}}\right)\nonumber\\
    &\qquad\qquad\qquad\qquad\times\mathbf{G}(\mathbf{r},\mathbf{r}',\omega)\cdot \opvec{b}(\mathbf{r}',\omega) +\mathrm{H.a.},
\end{align}
where we have used \(\epsilon(\mathbf{r},\omega)|_{r\in V_{\rm out}} = \epsilon_B\) to restrict the integration to the cavity volumes. Since the integration volume is finite, we can immediately set \(\epsilon_{B,I} = 0\) (cf. Ref.~\cite{franke2020fluctuation}).
Using the decomposition of the noise operators \(\hat{\mathbf{b}}(\mathbf{r},\omega)\) from Eq.~\eqref{eq:bdecomp}, we separate QNM and bath contributions to the incoming fields. The bath field is contained in the system-bath coupling [cf. Eq.~\eqref{eq:qnmham}]. In the QNM contribution, it follows from the definition of \(\mathbf{L}^*_{k_{\mu}}(\mathbf{r},\omega)\) [Eq.~\eqref{eq:Ldef}] that only contributions from the modes of cavity \(k\) survive, and hence
\begin{align}\label{appeq:evanescentfields}
    &\opvec{E}_{QNM}^{\mathrm{inc}}(\mathbf{r})\Big|_{r\in V_i} = i\sqrt{\frac{\hbar}{\pi\epsilon_0}} \sum_{k\neq i}\sum_{\mu\eta}\sqrt{\frac{2}{\pi\omega_{k_{\eta}}}}\int_0^{\infty}\mathrm{d}\omega A_{k_{\eta}}^*(\omega)\nonumber\\
    &\times\int_{V_k}\mathrm{d}^3r'\epsilon_I(\mathbf{r}',\omega)\mathbf{G}(\mathbf{r},\mathbf{r}',\omega)\qnm{f}{k_{\eta}}^*(\mathbf{r}')\left(S^{-1/2}\right)_{k_{\eta}k_{\mu}}\hat{a}_{k_{\mu}}\nonumber\\
    &\qquad\qquad\qquad\qquad\qquad\qquad\qquad\qquad+\mathrm{H.a.}
\end{align}

Here, the Green's function ensures proper retardation of the term. It is clear that, without sufficient time delay to match the spatial separation \(|\mathbf{r}-\mathbf{r}' | \approx R_{ik}\) between the cavities, the effect of the incoming fields will be small. An exact expansion of the Green's function in terms of the QNM fields is problematic since \(r\in V_i\) and \(r'\in V_k\). However, to understand the physical meaning of these contributions, it is instructive to perform an approximate expansion of the Green's function in terms of only the modes from cavity \(i\). We neglect inscattering from other cavities and obtain
\begin{align} \label{appeq:qnmincnrad}
    &\opvec{E}_{QNM}^{\mathrm{inc}}(\mathbf{r})\Big|_{r\in V_i} \nonumber\\
    &\approx i\sqrt{\frac{\hbar}{2\epsilon_0}}\sum_{k\neq i}\sum_{\lambda\eta\mu}\sqrt{\omega_{i_{\lambda}}}\qnm{f}{i_{\lambda}}(\mathbf{r})S^{\rm nrad}_{i_{\lambda}k_{\eta}}\left(S^{-1/2}\right)_{k_{\eta}k_{\mu}}\hat{a}_{k_{\mu}}\nonumber\\
    &\qquad\qquad\qquad\qquad\qquad\qquad\qquad\qquad\qquad+\mathrm{H.a.},
\end{align}
where 
\begin{align}  \label{eq:Snradinter}
    &S^{\rm nrad}_{i_{\lambda}k_{\eta}} = \frac{2}{\pi\sqrt{\omega_{i_{\lambda}}\omega_{k_{\eta}}}}\int_0^{\infty}\mathrm{d}\omega\, A_{i_{\lambda}}(\omega)A^*_{k_{\eta}}(\omega)\nonumber\\
    &\qquad\qquad\times\int_{V_k}\mathrm{d}^3r\epsilon_I(\mathbf{r},\omega)\qnm{F}{i_{\lambda}}(\mathbf{r},\omega)\cdot\qnm{f}{k_{\eta}}^*(\mathbf{r})
\end{align}
represents a contribution caused by the overlap between the modes \(i_{\lambda}\) and \(k_{\eta}\) due to absorption in cavity \(k\) [cf.~Eq.~\eqref{eq:Snraddef}], which would be a non-radiative contribution if the system were quantized as a whole. Equation~\eqref{appeq:qnmincnrad} can, therefore, be interpreted to represent the (instantaneous) contributions to the field inside cavity \(i\) due to non-radiative coupling to other cavities. 

Since the sources of \(\opvec{E}^{\rm inc}_{QNM}(\mathbf{r})\) lie in cavity \(k\), an alternative expansion of the Green's function in terms the modes of that cavity yields,
\begin{align} \label{appeq:qnmincprop}
    &\opvec{E}_{QNM}^{\mathrm{inc}}(\mathbf{r})\Big|_{r\in V_i}\approx i\sqrt{\frac{\hbar}{2\epsilon_0}}\sum_{k\neq i}\sum_{\lambda\eta\mu}\sqrt{\omega_{k_{\lambda}}}\nonumber\\
    &\times\int_0^{\infty}\mathrm{d}\omega \qnm{F}{k_{\lambda}}(\mathbf{r},\omega)S^{\rm nrad}_{k_{\lambda}k_{\eta}}(\omega)\left(S^{-1/2}\right)_{k_{\eta}k_{\mu}}\hat{a}_{k_{\mu}}+\mathrm{H.a.},
\end{align}
where \(S^{\rm nrad}_{k_{\lambda}k_{\eta}} = \int_0^{\infty}\mathrm{d}\omega S^{\rm nrad}_{k_{\lambda}k_{\eta}}(\omega)\) is the contribution to the intra-cavity overlap matrix from Eq.~\eqref{eq:Snraddef}, which would be included in the non-radiative part if the system were quantized as a whole. Equation~\eqref{appeq:qnmincprop} has the shape of incoming radiation to cavity \(i\) from other cavities. 

We stress that the expansions of the Green's function performed here are approximate. The full influence of the other cavities on the field inside cavity \(i\) contains contributions both from incoming radiation and non-radiative coupling between the cavities, as well as contributions due to higher order scattering processes (cf.~Appendix~\ref{appsec:greenfunc}). A full discussion of these terms requires a reliable method for expanding the Green's function in terms of QNMs of spatially far separated cavities, such as an extension on the existing QNM coupled mode theory \cite{ren2021quasinormal} including the regularized fields \(\qnm{F}{i_{\mu}}(\mathbf{r},\omega)\). The theory developed in the main text, however, assumes well-separated cavities, i.e., \(P_{ik}\gg 0\) [cf. Eq.~\eqref{eq:cavsepparam}], for which the influence of \(\opvec{E}^{\rm inc}_{QNM}(\mathbf{r})\) is assumed to be small.

\section{Inter-cavity overlap integral}\label{appsec:Smatrix}
After inserting the definition of the QNM operators \(\Tilde{\alpha}_{i_{\mu}}\) [given explicitly below Eq.~\eqref{eq:qnmefield}], the full overlap matrix \(\left[ \tilde{\alpha}_{i_{\mu}}, \tilde{\alpha}^{\dagger}_{j_{\eta}}\right] = S_{i_{\mu}j_{\eta}}\) reads,
\begin{align} \label{appeq:Smatrixdef}
    &S_{i_{\mu}j_{\eta}} = \frac{2}{\pi\sqrt{\omega_{i_{\mu}}\omega_{j_{\eta}}}}\int_0^{\infty}\mathrm{d}\omega\, A_{i_{\mu}}(\omega)A^*_{j_{\eta}}(\omega)\nonumber\\
    &\times\Big[\delta_{ij}\int_{V_i}\mathrm{d}^3r\epsilon_I(\mathbf{r},\omega)\qnm{f}{i_{\mu}}(\mathbf{r})\cdot\qnm{f}{j_{\eta}}^*(\mathbf{r})\nonumber\\
    &\quad +\int_{\overline{V_{ij}}}\mathrm{d}^3r\epsilon^{\alpha}_{B,I}\qnm{F}{i_{\mu}}(\mathbf{r},\omega)\cdot\qnm{F}{j_{\eta}}^*(\mathbf{r},\omega)\nonumber\\
    &\quad +(1-\delta_{ij})\int_{V_i}\mathrm{d}^3r\sqrt{\epsilon_I(\mathbf{r},\omega)}\sqrt{\epsilon^{\alpha}_{B,I}}\qnm{f}{i_{\mu}}(\mathbf{r})\cdot\qnm{F}{j_{\eta}}^*(\mathbf{r},\omega)\nonumber\\
    &\quad +(1-\delta_{ij})\int_{V_j}\mathrm{d}^3r\sqrt{\epsilon_I(\mathbf{r},\omega)}\sqrt{\epsilon^{\alpha}_{B,I}}\qnm{F}{i_{\mu}}(\mathbf{r},\omega)\cdot\qnm{f}{j_{\eta}}^*(\mathbf{r})\Big],
\end{align}
where \(\overline{V_{ij}} = \overline{V_i}\cap\overline{V_j}\) is the complement of the two cavity volumes. In the last two terms, we can perform the limit \(\lim_{\alpha\to 0}\epsilon^{\alpha}_B = \epsilon_B\) immediately since the integrals run over the finite cavity volumes. These contributions to the overlap matrix vanish in this limit since \(\epsilon_B\) is assumed to be real. However, the volume approaches infinity in the second term, so the integral will be evaluated first. 

\begin{figure}
    \centering
    \includegraphics[width=0.7\columnwidth]{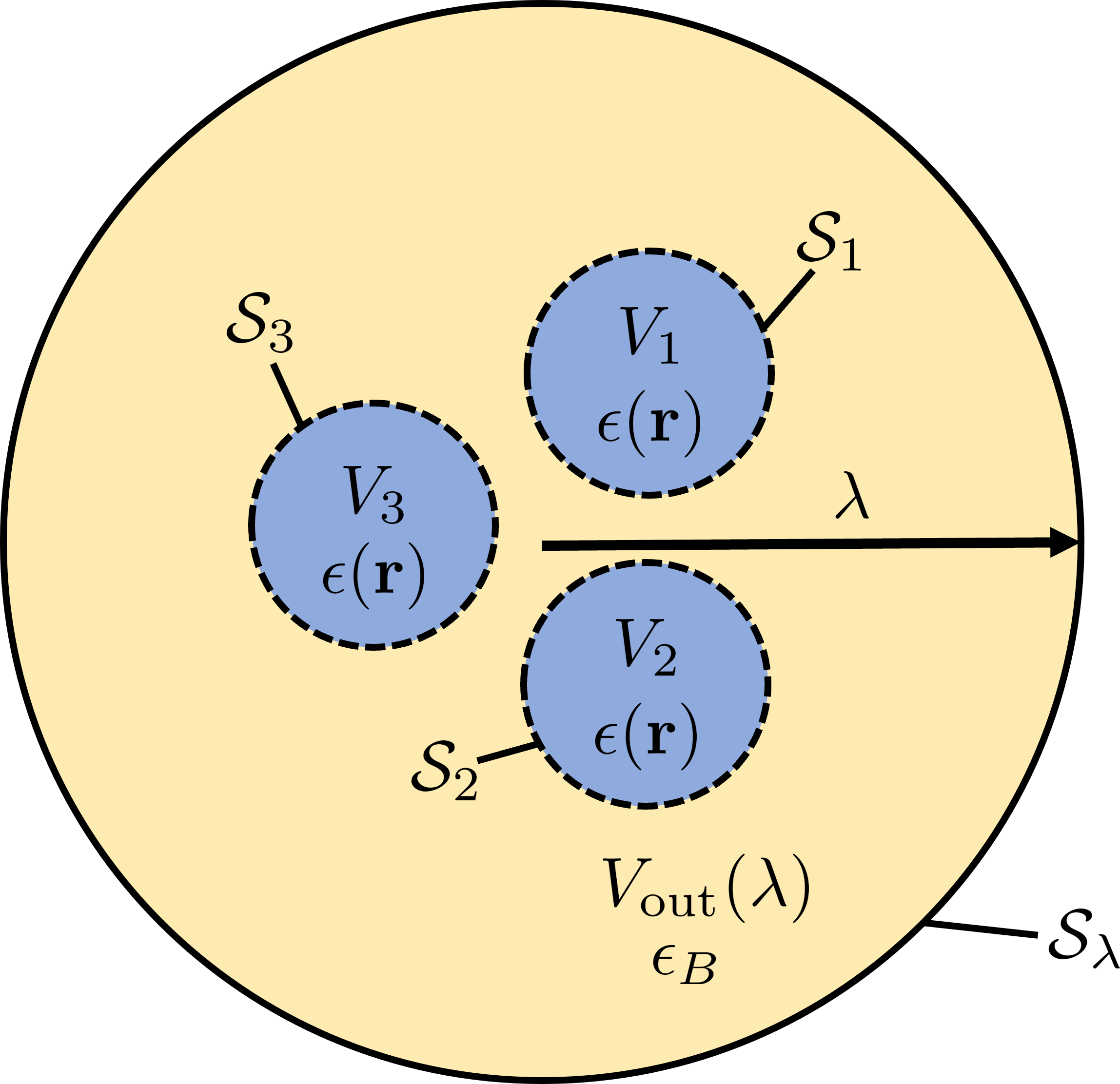}
    \caption{The spherical volume \(V_{\rm out}(\lambda)\) of radius \(\lambda\) encompasses all cavities. The entire space is recovered in the limit \(\lambda\to\infty\). \(\mathcal{S}_i\) and \(\mathcal{S}_{\lambda}\) are the surfaces of the volumes \(V_i\) and \(V_{\rm out}(\lambda)\), respectively.}
    \label{fig:V_out}
\end{figure}

Inserting the definition of the regularized QNMs \(\qnm{F}{i_{\mu}}\), the spatial integral outside the two cavities becomes
\begin{align*}
    &\lim_{\lambda\to\infty}\int_{\overline{V_{ij}}(\lambda)}\mathrm{d}^3r\epsilon^{\alpha}_{B,I}\qnm{F}{i_{\mu}}(\mathbf{r},\omega)\cdot\qnm{F}{j_{\eta}}^*(\mathbf{r},\omega) =\nonumber\\
    &\lim_{\lambda\to\infty}\int_{V_i}\mathrm{d}^3r'\int_{V_j}\mathrm{d}^3r''\qnm{f}{i_{\mu}}(\mathbf{r}')\Delta\epsilon(\mathbf{r}',\omega)\cdot \mathbf{M}_{\lambda,\alpha}(\mathbf{r}',\mathbf{r}'',\omega)\nonumber\\
    &\qquad\qquad\qquad\qquad\qquad\qquad\times\Delta\epsilon^*(\mathbf{r}'',\omega)\qnm{f}{j_{\eta}}^*(\mathbf{r}'',\omega),
\end{align*}
where
\begin{align*}
    &\mathbf{M}_{\lambda,\alpha}(\mathbf{r}',\mathbf{r}'',\omega)\nonumber\\
    &\qquad= \int_{\overline{V_{ij}}(\lambda)}\mathrm{d}^3r\epsilon^{\alpha}_{B,I}\mathbf{G}_B(\mathbf{r}',\mathbf{r},\omega)\cdot\mathbf{G}^*_B(\mathbf{r},\mathbf{r}'',\omega).
\end{align*}
Here, \(\overline{V_{ij}}(\lambda)\) is a spherical volume of radius \(\lambda\) including all cavities, with the cavity volumes \(V_i\) and \(V_j\) removed, so that \(\lim_{\lambda\to\infty}\overline{V_{ij}}(\lambda) = \overline{V_{ij}}\) (cf. Fig.~\ref{fig:V_out}). 

We apply the Helmholtz equation to the background Green's function [Eq.~\eqref{eq:helmgreen}] and use the dyadic-dyadic second Green's identity
\begin{align}\label{appeq:secondgreens}
    &\int_V\text{d}^3r\,\left[ (\nabla\times\nabla\times\mathbf{Q})^T\cdot\mathbf{P}-\mathbf{Q}^T\cdot(\nabla\times\nabla\times\mathbf{P})\right]\nonumber\\
    &=\oint_{\partial V}\text{d}A_s\left[(\opvec{n}_s\times\mathbf{Q})^T\cdot(\nabla\times\mathbf{P})-(\nabla\times\mathbf{Q})^T\cdot(\opvec{n}_s\times\mathbf{P})\right]
\end{align}
to obtain the surface integral
\begin{align*}
    &\frac{2i\omega^2}{c^2}\mathbf{M}_{\lambda,\alpha}(\mathbf{r},\mathbf{r}'',\omega) \nonumber\\
    &= \oint_{\mathcal{S}_i\cup\mathcal{S}_j}\mathrm{d}A_s \left\{[\opvec{n}_s\times\mathbf{G}_B(\mathbf{s},\mathbf{r}',\omega)]^T\cdot[\nabla_s\times\mathbf{G}^*_B(\mathbf{s},\mathbf{r}'',\omega)]\right.\nonumber\\
    &\qquad\qquad-\left.[\nabla_s\times\mathbf{G}_B(\mathbf{s},\mathbf{r}',\omega)]^T\cdot[\opvec{n}_s\times\mathbf{G}^*_B(\mathbf{s},\mathbf{r}'',\omega)]\right\}\nonumber\\
    &+\oint_{\mathcal{S}_{\lambda}}\mathrm{d}A_s \left\{[\opvec{n}_s\times\mathbf{G}_B(\mathbf{s},\mathbf{r}',\omega)]^T\cdot[\nabla_s\times\mathbf{G}^*_B(\mathbf{s},\mathbf{r}'',\omega)]\right.\nonumber\\
    &\qquad\qquad-\left.[\nabla_s\times\mathbf{G}_B(\mathbf{s},\mathbf{r}',\omega)]^T\cdot[\opvec{n}_s\times\mathbf{G}^*_B(\mathbf{s},\mathbf{r}'',\omega)]\right\},
\end{align*}
where \(\mathcal{S}_i\) is the surface of the cavity volume \(V_i\), while \(\mathcal{S}_{\lambda}\) is the outer surface shell of radius \(\lambda\) (Fig.~\ref{fig:V_out}), and \(\opvec{n}_s\) is the surface vector on the surface \(\mathcal{S}\), pointing into the volume. In the limit \(\lambda\to\infty\), the \(\mathcal{S}_{\lambda}\)-integral vanishes since the Green's functions obey the Silver-Müller radiation condition [cf. Eq~\eqref{eq:qnmsilvmüll}]. As there are no sources in the far field, the net flow of energy through a closed surface in the far field is zero due to energy conservation. 

After taking the limit \(\lambda\to\infty\), we also perform the limit \(\alpha\to 0\) (cf. Ref.~\onlinecite{franke2020fluctuation}). Furthermore, we insert the surface integrals back into Eq.~\eqref{appeq:Smatrixdef} to obtain the following expression for the QNM overlap integral
\begin{align} \label{appeq:Smatrixwithsurface}
    &S_{i_{\mu}j_{\eta}} = \frac{2}{\pi\sqrt{\omega_{i_{\mu}}\omega_{j_{\eta}}}}\int_0^{\infty}\mathrm{d}\omega\, A_{i_{\mu}}(\omega)A^*_{j_{\eta}}(\omega)\nonumber\\
    &\times\Big\{\delta_{ij}\int_{V_i}\mathrm{d}^3r\epsilon_I(\mathbf{r},\omega)\qnm{f}{i_{\mu}}(\mathbf{r})\cdot\qnm{f}{j_{\eta}}^*(\mathbf{r})\nonumber\\
    &+\frac{1}{2\omega\epsilon_0}\oint_{\mathcal{S}_i}\mathrm{d}A_s\Big[\left(\qnm{H}{i_{\mu}}(\mathbf{s},\omega)\times\opvec{n}_s\right)\cdot\qnm{F}{j_{\eta}}^*(\mathbf{s},\omega)\nonumber\\
    &\qquad\qquad\qquad\qquad\qquad+\mathrm{c.c.}(i_{\mu}\leftrightarrow j_{\eta})\Big]\nonumber\\
    &+\frac{(1-\delta_{ij})}{2\omega\epsilon_0}\oint_{\mathcal{S}_j}\mathrm{d}A_s\Big[\left(\qnm{H}{i_{\mu}}(\mathbf{s},\omega)\times\opvec{n}_s\right)\cdot\qnm{F}{j_{\eta}}^*(\mathbf{s},\omega)\nonumber\\
    &\qquad\qquad\qquad\qquad\qquad+\mathrm{c.c.}(i_{\mu}\leftrightarrow j_{\eta})\Big]\Big\},
\end{align}
where we have defined the regularized magnetic field QNMs \(\qnm{H}{i_{\mu}} = -i\nabla\times\qnm{F}{i_{\mu}}/(\omega\mu_0)\), and \(\opvec{n}_s\) now points out of the cavity volumes. The form of the overlap matrix in Eq.~\eqref{appeq:Smatrixwithsurface} greatly reduces the numerical cost since it only contains integrals over the (small) cavity volumes and cavity surfaces instead of the integrals over the entire space as in Eq.~\eqref{appeq:Smatrixdef}. We finally retrieve the expressions from Eqs.~\eqref{eq:Snraddef}, \eqref{eq:Sraddef}, and \eqref{eq:Sinterdef} for the two cases \(i=j\) and \(i\neq j\).

\section{Derivation of the cavity separation parameter}\label{appsec:cavsepparam}
\textit{Spatial separation.} To show the effect of the spatial separation on the equal-time overlap matrix \(S^{\rm inter}_{i_{\mu}j_{\eta}}\), we consider the integral
\begin{align}\label{appeq:examplint}
    &\mathcal{I}_{i_{\mu}j_{\eta}} =\int_0^{\infty}\mathrm{d}\omega\, \frac{\omega}{(\omega-\Tilde{\omega}_{i_{\mu}})(\omega-\Tilde{\omega}^*_{j_{\eta}})}\nonumber\\
    &\qquad\quad\times\oint_{\mathcal{S}_i}\mathrm{d}A_s\left[\qnm{H}{i_{\mu}}(\mathbf{s},\omega)\times\opvec{n}_s\right]\cdot\qnm{F}{j_{\eta}}^*(\mathbf{s},\omega),
\end{align}
which appears in the \(S^{\rm inter}\)-matrix (except for constant factors) from Eq.~\eqref{eq:Sinterdef}. 
Since \(\mathbf{s}\) lies on the surface of cavity \(i\) and we assume the distance \(R_{ij}\) between the cavities to be much larger than the size of the cavities, we neglect the propagation for \(\qnm{H}{i_{\mu}}\) and separate \(\qnm{F}{j_{\eta}}^{*}\) into a fast rotating exponential and an envelope function \(\qnm{F}{j_{\eta}}'^{*}\), as explained in the main text. Within a Fraunhofer approximation, we assume the envelope functions \(\qnm{F}{j_{\eta}}'^{*}\) to vary slowly with \(\omega\) compared to the exponential. This way, we apply the residue theorem to obtain
\begin{align*}
    &\mathcal{I}_{i_{\mu}j_{\eta}} = -2\pi \frac{\tilde{\omega}_{i_{\mu}}}{i(\Tilde{\omega}_{i_{\mu}}-\Tilde{\omega}^*_{j_{\eta}})}\mathrm{e}^{-i\Tilde{\omega}_{i_{\mu}}\tau_{ij}}\nonumber\\
    &\qquad\qquad\times\oint_{\mathcal{S}_i}\mathrm{d}A_s\tilde{\mathbf{J}}_{i_{\mu}}(\mathbf{s})\cdot\qnm{F}{j_{\eta}}'^*(\mathbf{s},\Tilde{\omega}_{i_{\mu}}),
\end{align*}
which scales as \(\mathrm{e}^{-\gamma_{i_{\mu}}\tau_{ij}}\), where \(\tau_{ij} = n_BR_{ij}/c\) is the retardation, and \(\tilde{\mathbf{J}}_{i_{\mu}}(\mathbf{s}) = \opvec{n}_s\times\qnm{H}{i_{\mu}}(\mathbf{s},\Tilde{\omega}_{i_{\mu}}) = \opvec{n}_s\times\qnm{h}{i_{\mu}}(\mathbf{s}) \) \cite{ren2020near} is the electric surface current. The inter-cavity overlap contains integrals like \(\mathcal{I}_{i_{\mu}j_{\eta}}\) over the surfaces of the two cavities. Since there is no scaling with distance in the intra-cavity overlap (\(R_{ii}=0\)), this yields, for the relative inter-cavity overlap, the scaling from Eq.~\eqref{eq:cavsepparam}.

\textit{Quality factor.} Apart from the spatial separation, the quality of the cavity modes \(Q_{i_{\mu}} = \omega_{i_{\mu}}/(2\gamma_{i_{\mu}})\) also plays a significant role in the relative magnitude of the inter-cavity overlap. Since a high-$Q$ cavity leaks energy at a slower rate, the overlap with other modes is also small. In the limit of a closed cavity, the overlap with modes of other cavities vanishes completely. 
In \cite{franke2020fluctuation}, it was shown that in the limit \(Q\to\infty\) (vanishing dissipation), the normal mode case is retained from the QNM theory, i.e., \(S_{i_{\mu}i_{\eta}}\to \delta_{\mu\eta}\). In this limit, \(S^{\rm nrad}_{i_{\mu}i_{\mu}}\to 0\), while (cf.~\cite{ren2020near, franke2020fluctuation})
\begin{align}
    &S^{\rm rad}_{i_{\mu}i_{\mu}}\approx \frac{1}{\epsilon_0i(\tilde{\omega}_{i_{\mu}}-\tilde{\omega}^*_{i_{\mu}})} \nonumber\\
    &\times\oint_{\mathcal{S}_i}\mathrm{d}A_s \opvec{n}_s\cdot\Big[\qnm{F}{i_{\mu}}(\mathbf{s},\omega_{i_{\mu}})\times\qnm{H}{i_{\mu}}^*(\mathbf{s},\omega_{i_{\mu}})+\mathrm{c.c.}\Big]\to 1.
\end{align}

Since the factor before the integral scales with \(Q_{i_{\mu}}\), it follows that the surface integral scales with \(1/Q_{i_{\mu}}\) so that the expression stays finite. However, the surface \(\mathcal{S}_i\) is not uniquely defined. Indeed, the surface of any closed volume that contains the resonator can be chosen for \(\mathcal{S}_i\) (for example, a spherical surface in the very far field is often chosen for numerical calculations \cite{ren2020near}, regardless of the specific resonator geometry). Hence, the scaling of the integral is independent of the specific shape of the surface and is instead a local property of the integrand. Since electric and magnetic fields are related via Maxwell's equations, we find the following scaling in the high-$Q$ limit,
\begin{align*}
    \qnm{F}{i_{\mu}}(\mathbf{r},\omega) \sim \frac{1}{\sqrt{Q_{i_{\mu}}}},
\end{align*}
and similarly for the magnetic field QNMs \(\qnm{H}{i_{\mu}}(\mathbf{r},\omega)\).
This scaling results from the open boundary conditions [Eq.~\eqref{eq:qnmsilvmüll}] becoming closed boundary conditions in the normal mode case, so the regularized fields outside the cavity vanish. 

We express this behavior of the overlap matrix via
\begin{align}
    S_{i_{\mu}j_{\eta}} = \delta_{ij}\delta_{\mu\eta}+\frac{1}{\sqrt{Q_{i_{\mu}}Q_{j_{\eta}}}}K_{i_{\mu}j_{\eta}},
\end{align}
where \(K_{i_{\mu}j_{\eta}}\) denotes the difference between the overlap matrix and the Kronecker delta, which is assumed to stay finite in the limit \(Q\to\infty\), so that the overlap matrix approaches a Kronecker delta with a scaling of \((Q_{i_{\mu}}Q_{j_{\eta}})^{-1/2}\), causing off-diagonal elements to vanish, while the diagonal elements stay finite.

\textit{Directionality and orientation.}
Consider the integral \(\mathcal{I}_{i_{\mu}j_{\eta}}\) from Eq.~\eqref{appeq:examplint}. Since the spatial integral runs over the surface of the \(i\)-th cavity, the scaling of the integral (and, by extension, the inter-cavity overlap) relies on the directionality of the emission from cavity \(j\). If cavity \(j\) emits its radiation away from cavity \(i\), then \(\qnm{F}{j_{\eta}}^*(\mathbf{s},\omega)\) will be zero on the surface of that cavity. 
From the theory of classical antennas, we obtain the definition for the directivity of an emission \cite{balanis2016antenna}
\begin{align*}
    \tilde{D}_{j_{\eta}}(\theta, \phi) = \frac{U_{j_{\eta}}(\theta,\phi)}{P_{j_{\eta}}^{\mathrm{tot}}/4\pi},
\end{align*}
where \(U_{j_{\eta}}(\theta,\phi)\) is the radiation intensity of the emission by mode \(j_{\eta}\) in the direction given by the angles \(\theta, \phi\), and \(P_{j_{\eta}}^{\mathrm{tot}}\) is the total radiated power of the emission. Let now \(\Omega_{ji}\) be the solid angle under which an emission from cavity \(j\) reaches cavity \(i\). By integrating the directivity over this angle and dividing by \(4\pi\), we obtain the fraction of the total power that is radiated into the solid angle \(\Omega_{ji}\)
\begin{align*}
    D_{j_{\eta}}(\Omega_{ji}) = \frac{P_{j_{\eta}}^{\Omega_{ji}}}{P_{j_{\eta}}^{\mathrm{tot}}} = \frac{\langle U_{j_{\eta}}\rangle_{\Omega_{ji}}}{\langle U_{j_{\eta}}\rangle_{4\pi}},
\end{align*}
where \(\langle U\rangle_{\Omega}\) denotes the average intensity over the solid angle \(\Omega\). Since the intra-cavity overlap is independent of the directionality of the emission and relative orientation, we obtain a scaling of the relative magnitude of the inter-cavity overlap with the directivity of the cavity emission in the direction of the other cavity.

The condition for a separate quantization of two cavities is that \(|S^{\rm inter}/S^{\rm intra}| \ll 1\). Hence, the overlap \(S_{i_{\mu}j_{\eta}}\) is negligible, if 
\begin{align*}
    &S_{i_{\mu}j_{\eta}}\sim\mathrm{e}^{-P_{i_{\mu}j_{\eta}}}\equiv\frac{\mathrm{e}^{-\gamma^{\min}_{i_{\mu}j_{\eta}}\tau_{ij}}}{\sqrt{Q_{i_{\mu}}Q_{j_{\eta}}}}D^{\max}_{i_{\mu}j_{\eta}} \ll 1\nonumber\\
    \Leftrightarrow & P_{i_{\mu}j_{\eta}} =\gamma^{\min}_{i_{\mu}j_{\eta}}\tau_{ij} +\mathrm{ln}\left(\sqrt{Q_{i_{\mu}}Q_{j_{\eta}}}\right)-\mathrm{ln}(D^{\max}_{i_{\mu}j_{\eta}}) \gg 0.
\end{align*}

\textit{Additional aspects.} In principle, the cavity separation parameter can be amended by additional contributions. For a geometry where the regularized QNMs \(\qnm{F}{i_{\mu}}(\mathbf{r},\omega)\) behave like spherical waves, their amplitude decreases with distance as \(1/r\), leading to a further decrease in the relative magnitude of the inter-cavity overlap. Since these other aspects are highly dependent on the system's dimensionality and specific geometry, we omit them here to leave the discussion general. We note that, in practice, a reasonable estimate of the cavity separation may be obtained even without an exact knowledge of all geometrical aspects. For example, for the overlap of the metal nanorods from Fig.~\ref{fig:overlapquant}(b), the directivity was approximated using a simple dipole model, which still yields good agreement with the full calculations, especially for larger separations. Thus, the cavity separation parameter is obtained from few basic properties of the QNMs without costly numerical calculations.

\section{Connection to phenomenological models}\label{appsec:pheno}
We show how the QNM Hamiltonian derived in Sec.~\ref{sec:qnmham} is connected to the phenomenological models (e.g., Ref.~\cite{carmichael1993quantum}) for sufficiently high-Q cavities. For this, we use the orthogonality relation
\begin{align} \label{appeq:cortho}
    \int\mathrm{d}^3r\int_0^{\infty}\mathrm{d}\omega \mathbf{L}_{i_{\mu}}(\mathbf{r},\omega)\cdot \opvec{c}(\mathbf{r},\omega) = 0,
\end{align}
for all \(i_{\mu}\), to rewrite the system-bath coupling Hamiltonian to read
\begin{align} 
    H_{SB} = \hbar\sum_{i,\mu}\int\mathrm{d}^3r\int_0^{\infty}\mathrm{d}\omega \mathbf{g}_{i_{\mu}}(\mathbf{r},\omega)\cdot \opvec{c}(\mathbf{r},\omega)\hat{a}^{\dagger}_{i_{\mu}}+\mathrm{H.a.},
\end{align}
where 
\begin{align} \label{appeq:gdef}
    &\mathbf{g}_{i_{\mu}}(\mathbf{r},\omega) \nonumber\\
    &\qquad= \sum_{\mu'\mu''}(S^{-1/2})_{i_{\mu}i_{\mu'}}(\omega-\tilde{\omega}_{i_{\mu'}})(S^{1/2})_{i_{\mu'}i_{\mu''}}\mathbf{L}_{i_{\mu''}}(\mathbf{r},\omega)
\end{align}
is a QNM-bath coupling element. Next, we define new operators 
\begin{align}
    \Tilde{B}_{i_{\mu}} = \int\mathrm{d}^3r\int_0^{\infty}\mathrm{d}\omega \mathbf{g}_{i_{\mu}}(\mathbf{r},\omega)\cdot \opvec{c}(\mathbf{r},\omega).
\end{align}
When compared to the QNM projectors \(\mathbf{L}_{i_{\mu}}(\mathbf{r},\omega)\), the coupling elements \(\mathbf{g}_{i_{\mu}}(\mathbf{r},\omega)\) lack a pole at the QNM eigenfrequency and are thus associated with a broad range of frequencies instead of a single resonance. Hence, the new operators represent the quantum noise in the bath due to the cavities' emission or absorption of photons (cf. Refs.~\cite{gardiner1985input, oppermann2018quantum}). Note that while the bath operators \(\opvec{c}(\mathbf{r},\omega)\) are generally non-bosonic, we use here a bosonic approximation on the assumption that the QNMs are mostly confined to their respective cavities \cite{franke2020quantized}. This assumption holds well for high-Q cavities. Thus, we obtain the following commutator for the noise operators \(\Tilde{B}_{i_{\mu}}\):
\begin{align}
    &\left[\Tilde{B}_{i_{\mu}}(t),\Tilde{B}_{i_{\eta}}^{\dagger}(t')\right]\nonumber\\
    &\qquad= \int\mathrm{d}^3r\int_0^{\infty}\mathrm{d}\omega \mathbf{g}_{i_{\mu}}(\mathbf{r},\omega)\cdot \mathbf{g}_{i_{\eta}}(\mathbf{r},\omega) \mathrm{e}^{-i\omega(t-t')}.
\end{align}
The same relation was found in Ref.~\cite{franke2020quantized} in the context of deriving a quantum Langevin equation, and we use the Born-Markov approximation performed therein to obtain
\begin{align} \label{appeq:unsymmnoisecomm}
    &\left[\Tilde{B}_{i_{\mu}}(t),\Tilde{B}_{i_{\eta}}^{\dagger}(t')\right] = 2\chi^{(-)}_{i_{\mu}i_{\eta}}\delta(t-t'),
\end{align}
where \(\chi^{(-)}_{i_{\mu}i_{\eta}} = (i/2)\sum_{\mu'\eta'}(S^{-1/2})_{i_{\mu}i_{\mu'}}(\tilde{\omega}_{i_{\mu'}}-\tilde{\omega}^*_{i_{\eta'}}) \times S_{i_{\mu'}i_{\eta'}}(S^{-1/2})_{i_{\eta'}i_{\eta}}\) is the QNM dissipator matrix that also appears in the QNM-Jaynes-Cummings model that was derived in \cite{franke2020quantized} and is related to the dissipation of energy from the open cavity into the bath. From Eq.~\eqref{appeq:unsymmnoisecomm}, it is apparent that the noise created by emission/absorption of the \(i\)-th cavity is Gaussian in nature. To obtain bosonic noise operators, we perform the transformation
\begin{align}
    \hat{B}_{i_{\mu}} = -i\frac{1}{\sqrt{2}}\left[\bm{\chi}^{(-)}\right]^{-1/2}_{i_{\mu}i_{\eta}}\Tilde{B}_{i_{\eta}},
\end{align}
which yields the commutation relation \([\hat{B}_{i_{\mu}}(t),\hat{B}^{\dagger}_{i_{\eta}}(t')] = \delta_{\mu\eta}\delta(t-t')\). With these symmetrized noise operators, the system-bath Hamiltonian transforms into
\begin{align}\label{appeq:phenham}
    H_{SB} &= i\hbar\sum_{i,\mu\eta}\sqrt{2}\left[\bm{\chi}^{(-)}\right]^{1/2}_{i_{\mu}i_{\eta}}\hat{B}_{i_{\eta}}\hat{a}^{\dagger}_{i_{\mu}} +\mathrm{H.a.}\nonumber\\
    &\approx i\hbar \sum_{i,\mu}\sqrt{2\gamma_{i_{\mu}}}\hat{a}^{\dagger}_{i_{\mu}}\hat{B}_{i_{\mu}}+\mathrm{H.a.}
\end{align}
For the last line, we assumed that the off-diagonal coupling terms are negligible \cite{franke2020fluctuation}. Equation~\eqref{appeq:phenham} has precisely the shape of the phenomenological system-bath coupling from Ref.~\cite{carmichael1993quantum}.

\section{Pole approximation of the inter-cavity overlap}\label{appsec:poleapprox}
The overlap-integral from Eq.~\eqref{eq:Sinterdef} is demanding to calculate numerically since the surface integral must be computed separately for each frequency point. Instead, we present an efficient pole approximation that eliminates the frequency integration. For notational simplicity, we consider the overlap between two modes of separate cavities 1 and 2 ($i_{\mu}\rightarrow 1$, ~$j_{\eta}\rightarrow 2$, ~$\mathcal{S}_{i}\rightarrow \mathcal{S}_{1}$,~$\mathcal{S}_{j}\rightarrow \mathcal{S}_{2}$). The extension to the multi-mode case is straightforward. Thus, Eq.~\eqref{eq:Sinterdef} reads,
\begin{align} \label{eq:Sinterdef_fullw}
    &S^{\rm inter}_{12} =  \frac{2}{\pi\sqrt{\omega_{1}\omega_{2}}}\int_0^{\infty}\mathrm{d}\omega\,\frac{A_{1}(\omega)A^*_{2}(\omega)}{2\omega \epsilon_0}\nonumber\\
    \times &\Bigg\{\oint_{\mathcal{S}_{1}}\mathrm{d}A_{\mathcal{S}_{1}}\Big[\left(\qnm{H}{1}(\mathbf{r}^{\prime}_1,\omega)\times\opvec{n}_{1}\right)\cdot\qnm{F}{2}^*(\mathbf{r}^{\prime}_1+\mathbf{R}_{21},\omega)\nonumber\\
    &\qquad\qquad\qquad\qquad\qquad\qquad+\mathrm{c.c.}(1\leftrightarrow 2)\Big]\nonumber\\
    &+\oint_{\mathcal{S}_{2}}\mathrm{d}A_{\mathcal{S}_{2}}\Big[\left(\qnm{H}{1}(\mathbf{r}^{\prime}_2+\mathbf{R}_{12},\omega)\times\opvec{n}_{2}\right)\cdot\qnm{F}{2}^*(\mathbf{r}^{\prime}_2,\omega)\nonumber\\
    &\qquad\qquad\qquad\qquad\qquad\qquad+\mathrm{c.c.}(1\leftrightarrow 2)\Big]\Bigg\}
\end{align}
where \(\mathbf{r}^{\prime}_{1/2}\) runs from the center of cavity 1 (cavity 2) to the surface \(\mathcal{S}_1\) (\(\mathcal{S}_2\)). The regularized fields from the other cavity are appropriately shifted by the connecting vectors \(\mathbf{R}_{12}\) and \(\mathbf{R}_{21} = -\mathbf{R}_{12}\). Furthermore, \(\opvec{n}_{1/2}\) is the surface normal vector on \(\mathcal{S}_{1/2}\), pointing outwards.

Due to the poles contained in \(A_{1}(\omega)A^*_{2}(\omega)\), the frequency integral is naturally confined to a small frequency regime around the real part of the QNM eigenfrequencies \(\omega_{1/2}\). Thus, choosing an appropriate upper limit of the integral \(\omega_{\rm max} > \omega_{1/2}\) is sufficient for calculating the overlap. We discuss the results of this calculation below in comparison to the pole approximation.

To apply a (resonant) pole approximation, we assume that the distance \(|\mathbf{R}_{12}|\) between the cavities is large enough so that $\tilde{\mathbf{F}}_{1},~\tilde{\mathbf{H}}_{1}$ on \(\mathcal{S}_2\) ($\tilde{\mathbf{F}}^*_{2},~\tilde{\mathbf{H}}^*_{2}$ on \(\mathcal{S}_1\)) behave like outgoing waves with phase \(\mathrm{e}^{ikr}\) (\(\mathrm{e}^{-ikr}\)), where \(k=n_B\omega/c\). Then, for an inter-cavity distance that is large compared to the size of the cavities \(|\mathbf{R}_{12}|\gg |\mathbf{r}^{\prime}_{1/2}|\), the frequency integral is additionally dominated by a fast-rotating exponential \(\mathrm{e}^{ik|\mathbf{R}_{12}|}\) from the \(\mathcal{S}_2\)-integral (\(\mathrm{e}^{-ik|\mathbf{R}_{12}|}\) from the \(\mathcal{S}_1\)-integral). In contrast, all other contributions contained in the surface integral vary slowly with \(\omega\). We then extend the lower limit to \(-\infty\) and (approximately) solve the frequency integral using the residue theorem (cf.~Appendix~\ref{appsec:cavsepparam}), where the pole in the upper or lower half of the complex plane is chosen depending on the sign of the exponential.
We obtain,

\begin{align} \label{eq:Sinterdef_pole}
    &S^{\rm inter}_{\rm 12,pole} \nonumber\\
    &= -\frac{1}{2\epsilon_0}\frac{\tilde{\omega}_1}{\sqrt{\omega_{1}\omega_{2}}}\frac{1}{i(\tilde{\omega}_{1}-\tilde{\omega}_{2}^{*})}\oint_{\mathcal{S}_{1}}\mathrm{d}A_{\mathcal{S}_{1}}\Big[\qnm{
    J}{1}(\mathbf{r}_{1}^{\prime})\cdot\qnm{F}{2}^*(\mathbf{a}_{21},\tilde{\omega}_1)\nonumber\\
    &\qquad\qquad\qquad\qquad\qquad+\qnm{M}{1}(\mathbf{r}_{1}^{\prime})\cdot\qnm{H}{2}^*(\mathbf{a}_{21},\tilde{\omega}_1)\Big]\nonumber\\
    &-\frac{1}{2\epsilon_0}\frac{\tilde{\omega}_2^{*}}{\sqrt{\omega_{1}\omega_{2}}}\frac{1}{i(\tilde{\omega}_{1}-\tilde{\omega}_{2}^{*})}\oint_{\mathcal{S}_{2}}\mathrm{d}A_{\mathcal{S}_{2}}\Big[\qnm{M}{2}^*(\mathbf{r}_{2}^{\prime})\cdot\qnm{H}{1}(\mathbf{a}_{12},\tilde{\omega}_2^{*})\nonumber\\
    &\qquad\qquad\qquad\qquad\qquad+\qnm{J}{2}^*(\mathbf{r}_{2}^{\prime})\cdot\qnm{F}{1}(\mathbf{a}_{12},\tilde{\omega}_2^{*})\Big],
\end{align}
where \(\mathbf{a}_{12} = \mathbf{r}^{\prime}_{2}+\mathbf{R}_{12}\), \(\mathbf{a}_{21} = \mathbf{r}^{\prime}_{1}+\mathbf{R}_{21}\), and $\tilde{\mathbf{F}}_{2}^{*}(\mathbf{a}_{21},\tilde{\omega}_{1})=\left[\tilde{\mathbf{F}}_{2}^{*}(\mathbf{a}_{21},\omega)\right]\Big|_{\omega=\tilde{\omega}_1}=\left\{\left[\tilde{\mathbf{F}}_{2}(\mathbf{a}_{21},\omega)\right]^{*}\right\}\Big|_{\omega=\tilde{\omega}_1}$.

Moreover, we used \(\qnm{F}{1}(\mathbf{r}^{\prime}_1,\tilde{\omega}_1)=\qnm{f}{1}(\mathbf{r}^{\prime}_1)\), \(\qnm{H}{1}(\mathbf{r}^{\prime}_1,\tilde{\omega}_1)=\qnm{h}{1}(\mathbf{r}^{\prime}_1)\) on \(\mathcal{S}_1\) (and accordingly for \(\qnm{F}{2}^*,\qnm{H}{2}^*\) on \(\mathcal{S}_2\)) to obtain the electric (magnetic) surface currents on the near field surface $\mathcal{S}_{1/2}$~\cite{Neartofar_1992},
\begin{align}
\tilde{\mathbf{J}}_{\rm 1/2}(\mathbf{r}^{\prime}_{1/2})&=\mathbf{\hat{n}}_{1/2}\times\tilde{\mathbf{h}}_{1/2}(\mathbf{r}^{\prime}_{1/2}),\\
\tilde{\mathbf{M}}_{\rm 1/2}(\mathbf{r}^{\prime}_{1/2})&=-\mathbf{\hat{n}}_{1/2}\times\tilde{\mathbf{f}}_{1/2}(\mathbf{r}^{\prime}_{1/2}).
\end{align}
The magnetic QNMs are defined via
\begin{equation}\label{eq:Mag_QNM}
    \mathbf{\tilde{h}}_{1/2}(\mathbf{r}^{\prime}_{1/2})=\frac{1}{i\tilde{\omega}_{1/2}\mu_{0}}\nabla\times\mathbf{\tilde{f}}_{1/2}(\mathbf{r}^{\prime}_{1/2}).
\end{equation}

The dot products of surface currents and regularized QNMs are required for $S_{\rm 12}^{\rm inter}$. One can directly get the bare QNMs $\tilde{\mathbf{f}}_{1/2}$ [and $\tilde{\mathbf{h}}_{1/2}$ from Eq.~\eqref{eq:Mag_QNM}], and thus the surface currents ($\tilde{\mathbf{J}}_{\rm 1/2}$,~$\tilde{\mathbf{M}}_{\rm 1/2}$) at the near field surfaces $\mathcal{S}_{1/2}$. The required regularized QNMs $\tilde{\mathbf{F}}_{1/2},~\tilde{\mathbf{H}}_{1/2}$, are calculated using an efficient near field to far field (NF2FF) transformation~\cite{ren2020near}.

In  Fig.~\ref{fig:s12_Full_vs_pole} (b), we show the overlap $|S_{12}^{\rm inter}|$ for the example of two gold nanorods with distance $L_{\rm gap}=2000~$nm  from full frequency integration and pole approximation, which exhibit excellent agreement for sufficiently large \(\omega_{\rm max}\). We also see excellent agreements for other separations (not shown here).

\section{Simulation details for the metal nanorods}\label{appsec:simulationdetail}

\subsection{Model set-up}
For the 3D mode example in the main text, we have used a metal dimer with various gap separations in free space (i.e., with permittivity $\epsilon_{\rm B}=n_{\rm B}^{2}=1.0$). As shown in the inset of Fig.~\ref{fig:overlapquant} (b) [see also Fig.~\ref{fig:bareQNM}(a,b)], the length of the cylindrical metal nanorod 1 (2) is $L_1=90~$nm~($L_2=100~$nm) (and their longitudinal axes coincide). Both nanorods have the same diameter: $D_1=D_2=20~$nm. 

We use a  Drude model to describe the dielectric function of the gold-like nanorods
\begin{equation}\label{eq:Drude}
    \epsilon_{\rm M}=1-\frac{\omega_{\rm p}^{2}}{\omega^{2}+i\omega\gamma_{\rm p}},
\end{equation}
where $\hbar\omega_{\rm p}=8.2934$ eV, with  $\hbar\gamma_{\rm p}=0.0928$ eV for nanorod 1 and
$\hbar\gamma_{\rm p}=(1/3)\times0.0928$ eV
for nanorod 2.

To obtain the QNMs of the bare and coupled system (see Fig.~\ref{fig:dimer}), 
we employ a dipole excitation technique~\cite{bai_efficient_2013-1}, in complex frequency space, using a finite-element solver implemented in COMSOL. The total volume of the computational domain [including perfectly matched layers (PMLs)] is 
around $37$ $\mu$m$^{3}$. The total thickness of the PMLs is $600~$nm, which was found to efficiently minimize boundary reflections. We have used high-resolution mesh settings around the dipole excitation point, where the maximum element size is $0.1$ nm at the dipole point and within a small sphere with a radius of $1~$nm that encloses the dipole point. The maximum mesh element sizes inside and outside the metal nanorods are $2$ nm and $80~$nm, respectively.

\subsection{Bare QNM and coupled QNMs}

\begin{figure}
    \centering
    \includegraphics[width = 0.99\linewidth]{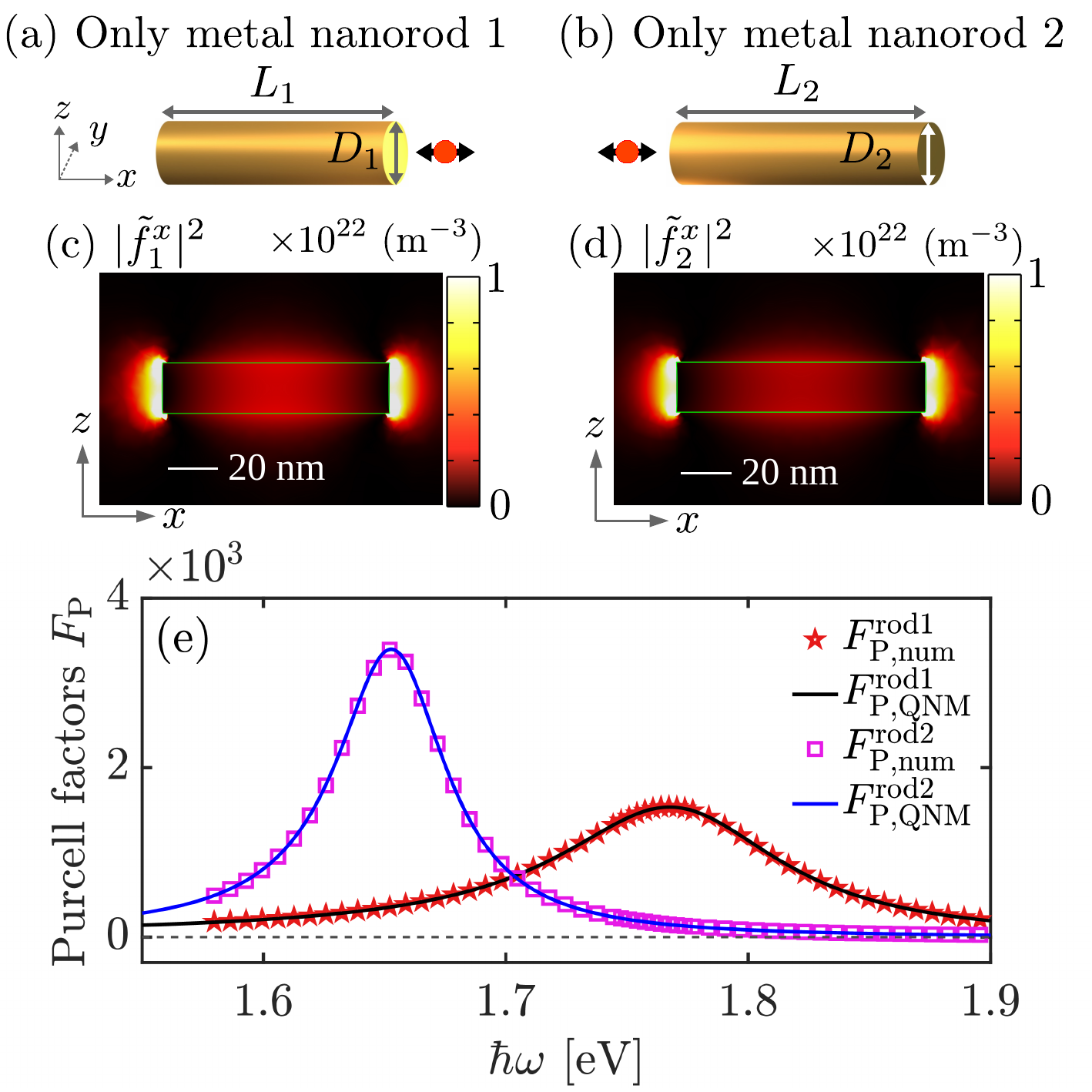}
    \caption{QNMs for isolated nanorods. (a) The schematic diagram for metal cylindrical nanorod 1 only, which is with a length of $L_{1}=90~$nm and diameter of $D_{1}=20~$nm. The  dipole point is placed at the longitudinal axis of the nanorod and is $10~$nm away from one end of the nanorod. (b) The schematic diagram for metal cylindrical nanorod 2 only, which is with a length of $L_{2}=100~$nm and diameter of $D_{2}=20~$nm. The dipole is placed $10~$nm away from the nanorod 2. (c) Mode profile $|\tilde{f}_{1}^{x}|^2$ for the dominant single QNM of metal nanorod 1 in the frequency regime of interest. (d) Mode profile $|\tilde{f}_{2}^{x}|^2$ for the dominant single QNM of metal nanorod 2. (e) The Purcell factors from QNMs results (solid curves) and full numerical dipole simulations (markers) for the dipole $10~$nm away from the isolated nanorods 1 and 2.
    }
    \label{fig:bareQNM}
\end{figure}

\begin{table*}[htb]
\caption { Eigenfrequencies and quality factors of coupled QNMs for dimer with various gap separations. The parameters for the bare QNM of metal nanorod 1 (nanorod 2) alone are also shown for comparison. 
} \label{table: cQNM} 
    \centering
    \begin{tabular}{|c|c|c|c|c|}
 \hline
Gap distance & $\hbar\tilde{\omega}_{\rm cQNM1}$ & $Q_{\rm cQNM1}$  & $\hbar\tilde{\omega}_{\rm cQNM2}$ & $Q_{\rm cQNM2}$\\
 \hline
$L_{\rm gap}=50$~nm & $(1.7847 - 0.0447i)~$eV & $20.0$  & $(1.6375 - 0.0382i)~$eV & $21.4$\\
 \hline
$L_{\rm gap}=100$~nm & $(1.7713 - 0.0525i)~$eV & $16.9$  & $(1.6523 - 0.0315i)~$eV & $26.2$\\
 \hline
$L_{\rm gap}=200$~nm & $(1.7699 - 0.0565i)~$eV & $15.7$  & $(1.6541 - 0.0280i)~$eV & $29.5$\\
 \hline
$L_{\rm gap}=500$~nm & $(1.7706 - 0.0566i)~$eV & $15.6$  & $(1.6536 - 0.0278i)~$eV & $29.7$\\
 \hline
$L_{\rm gap}=2000$~nm & $(1.7705 - 0.0566i)~$eV & $15.6$  & $(1.6536 - 0.0277i)~$eV & $29.8$\\
 \hline
Only metal nanorod 1 & $(1.7705-0.0566i)$ eV & $15.6$  & & \\
 \hline
Only metal nanorod 2 &  &  & $(1.6536-0.0277i)$ eV & $29.8$\\
 \hline

 \end{tabular}
\end{table*}

First, we investigate the individual QNMs for two metal nanorods. In the frequency regime of interest, there is a dominant single-QNM $\tilde{\mathbf{f}}_1$ for metal nanorod 1 only [Fig.~\ref{fig:bareQNM} (a)]. 
The eigenfrequency is $\hbar\tilde{\omega}_{1}=\hbar\omega_1-i\hbar\gamma_{1}=(1.7705-0.0566i)$ eV [Quality factor $Q_1=\omega_1/(2\gamma_1)\approx 15.6$, resonance wavelength $\lambda_1={2\pi c}/{\omega_1}\approx700~{\rm nm}$]. The field distribution ($|\tilde{f}_1^x|^2$, dominant component) is shown in Fig.~\ref{fig:bareQNM} (c). 

With the single mode approximation, the photon Green function is reconstructed as ~\cite{leung1994completeness,ge2014quasinormal},
\begin{align}\label{eq:GF}
\begin{split}
\mathbf{G}_{\rm rod1}\left(\mathbf{r},\mathbf{r}_{0},\omega\right)\approx A_{\rm 1}(\omega)\tilde{\mathbf{f}}_{\rm 1}\left({\bf r}\right)\tilde{\mathbf{f}}_{\rm 1}\left({\bf r}_{0}\right),
\end{split}
\end{align}
where $A_{\rm 1}(\omega)=\omega/[2(\tilde{\omega}_{\rm 1}-\omega)]$.

An $x$-polarized dipole (with a real dipole moment $\mathbf{d}=\mathbf{n}_{\rm d}|\mathbf{d}|$) is placed at $\mathbf{r}_0$, which is $10~$nm away from one end of the metal nanorod 1 [see Fig.~\ref{fig:bareQNM} (a)], the Purcell factors are shown in Fig.~\ref{fig:bareQNM}(e).  The solid black curve represents QNM results~($F_{\rm P,QNM}^{\rm rod1}$)~\cite{kristensen_modes_2014}
 \begin{align}
 \begin{split}\label{eq:Gamma_QNM}
 \Gamma^{\rm rod1}(\mathbf{r}_{0},\omega)&=\frac{2}{\hbar\epsilon_{0}}\mathbf{d}\cdot{\rm Im}\{\mathbf{G}_{\rm rod1}(\mathbf{r}_{0},\mathbf{r}_{0},\omega)\}\cdot\mathbf{d},\\
 F_{\rm P,QNM}^{\rm rod1}(\mathbf{r}_{0},\omega)&=\frac{\Gamma^{\rm rod1}(\mathbf{r}_{0},\omega)}{\Gamma_0(\omega)},\\
 \end{split}  
 \end{align}
where the homogeneous decay rate $\Gamma_0(\omega)=\frac{2}{\hbar\epsilon_{0}}\mathbf{d}\cdot{\rm Im}\{\mathbf{G}_{\rm B}(\mathbf{r}_{0},\mathbf{r}_{0},\omega)\}\cdot\mathbf{d}$ is related to the background Green function, where Im$\{\mathbf{G}_{\rm B}({\bf r}_0,{\bf r}_0,\omega)\}=(\omega^3n_{\rm B}/6\pi c^3)\mathbb{1}$. 

The red markers in Fig.~\ref{fig:bareQNM} (e) show the full dipole results ($F_{\rm P, num}^{\rm rod1}$),
\begin{equation}\label{eq:Purcellfulldipole}
    F_{\rm P,num}^{\rm rod1/rod2}(\mathbf{r}_{0},\omega)=\frac{\int_{ \Sigma_{\rm d}}\hat{\mathbf{n}}\cdot {\bf S}_{\rm }(\mathbf{r},\omega)d\mathbf{r} }{\int_{ \Sigma_{\rm d}}\hat{\mathbf{n}}\cdot {\bf S}_{\rm 0}(\mathbf{r},\omega)d\mathbf{r} }
\end{equation}
where $\Sigma_{\rm d}$ is a closed spherical (radius of $1$~nm) surface enclosing the dipole, and  ${\bf S}$ ($\mathbf{S}_0$) represent the Poynting vector with (without) the resonator.  

The excellent agreement between QNM results (black curve, $F_{\rm P,QNM}^{\rm rod1}$) and full numerical results (red markers, $F_{\rm P,num}^{\rm rod1}$) in Fig.~\ref{fig:bareQNM}(e) demonstrates the validity of the single QNM approximation for metal nanorod 1. 

For metal nanorod 2 only, the dominant single QNM $\tilde{\mathbf{f}}_2$ in the frequency regime of interest has eigenfrequency $\hbar\tilde{\omega}_{2}=\hbar\omega_2-i\hbar\gamma_{2}=(1.6536-0.0277i)$ eV [Quality factor $Q_2=\omega_2/(2\gamma_2)\approx 29.8$, resonance wavelength $\lambda_2={2\pi c}/{\omega_2}\approx750~{\rm nm}$]. The field distribution ($|\tilde{f}_{2}^{x}|^2$) is shown in Fig.~\ref{fig:bareQNM} (d). For an x-polarized dipole placed $10~$ nm away from one end of the metal nanorod 2 (see Fig.~\ref{fig:bareQNM} (b)), the Purcell factors are shown in Fig.~\ref{fig:bareQNM} (e). The solid blue curve represents $F_{\rm P,QNM}^{\rm rod2}$ from QNM results. Similar formulas are used as nanorod 1 but replacing $\mathbf{G}_{\rm rod1}$ by $\mathbf{G}_{\rm rod2}\left(\mathbf{r},\mathbf{r}_{0},\omega\right)\approx A_{\rm 2}(\omega)\tilde{\mathbf{f}}_{\rm 2}\left({\bf r}\right)\tilde{\mathbf{f}}_{\rm 2}\left({\bf r}_{0}\right)$ with $A_{\rm 2}(\omega)=\omega/[2(\tilde{\omega}_{\rm 2}-\omega)]$.
The magenta markers show the full dipole results from Eq.~\eqref{eq:Purcellfulldipole}, which also agrees perfectly with the single QNM approximation for nanorod 2.

To check the coupling effect, dimer structures consisting of these two nanorods are investigated. There are two coupled QNMs (cQNM 1 and cQNM 2) in the frequency regime of interest. The eigenfrequencies and quality factors with various separations are shown in Table.~\ref{table: cQNM}. The eigenfrequencies and quality factors for the bare QNMs of the individual nanorods are also compared. For gap distances $L_{\rm gap}=50,100,200~$nm, the field distribution $|\tilde{f}^{x}_{\rm cQNM1/2}|^2$ are shown in Fig.~\ref{fig:dimer}. For larger separations, coupled QNMs gradually approach the isolated QNM in terms of the eigenfrequencies, quality factors, and mode profile.
Focusing on separations \(L_{\rm gap}\geq 200\, {\rm nm}\) [Fig.~\ref{fig:overlapquant} (b)], we thus use the bare QNMs of the individual nanorods for all different gap distances.

\begin{figure*}
    \centering
    \includegraphics[width = 0.94\linewidth]{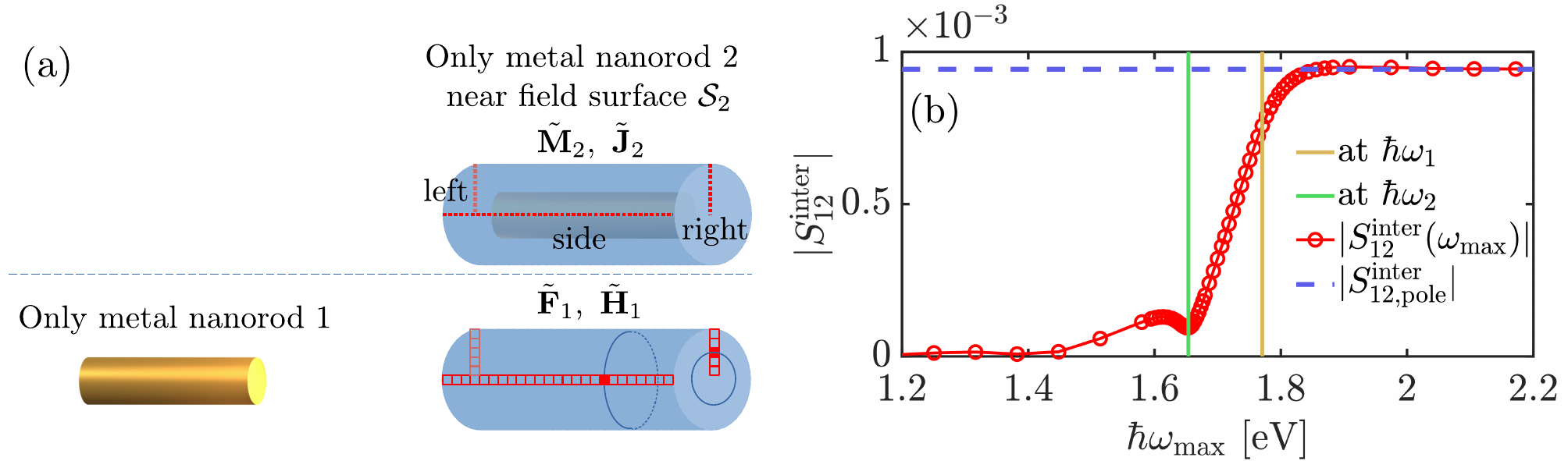}
    \caption{(a) Schematic diagram for the integral over the near surface $\mathbf{S}_{2}$ of nanorod 2. Firstly, to employ the simplification based on the axial symmetry of the radiation from the cylindrical metal nanorods to the longitudinal axis of the nanorods, $\mathbf{S}_{2}$ is selected as a cylindrical surface, which is from $50~$nm away from the surface of nanorod 2. As a result, one only has to get regularized QNMs ($\mathbf{\tilde{F}}_{1}$, $\mathbf{\tilde{H}}_{1}$) and surface currents ($\mathbf{\tilde{M}}_{2}$, $\mathbf{\tilde{J}}_2$) at three lines on surface $\mathbf{S}_2$, including a line at the side cylindrical surface, a line at the left circular surface and a line at the right circular surface. Secondly, different grid sizes are used for regularized QNMs ($5~$nm, relatively coarse grid) and surface currents ($0.5~$nm, relatively fine grid).
    (b) $|S_{12}^{\rm inter}|$ for $L_{\rm gap}=2000~$nm from full frequency integral (Eq.~\eqref{eq:Sinterdef_fullw}, red curve with markers) and pole approximation (Eq.~\eqref{eq:Sinterdef_pole}, dashed blue line). In total, 94 frequency points are used between $\hbar\omega\approx 0.1~$eV and $\hbar\omega\approx 3.3~$eV. Finer resolution between points is used close to the resonance $\hbar\omega_{2}$ (see vertical green line) and $\hbar\omega_1$ (see vertical orange line) of metal nanorod 2 and nanorod 1. When enough frequency ranges are included in the integral, the results $|S_{12}^{\rm inter}(\omega_{\max})|$ from a full frequency integration recovers $|S_{\rm 12,pole}^{\rm inter}|$ from a pole approximation. 
    }
    \label{fig:s12_Full_vs_pole}
\end{figure*}

\subsection{Simplifications of the surface integral}
To obtain the overlap from Eq.~\eqref{eq:Sinterdef_fullw} or Eq.~\eqref{eq:Sinterdef_pole}, the integration of the dot products of regularized QNMs and surface currents over surfaces are required. For example, as shown in Fig.~\ref{fig:s12_Full_vs_pole} (a), the near field surface $\mathcal{S}_2$ is selected as a cylindrical surface, which is $50~$nm away from the surface of nanorod 2. The surface currents $\mathbf{\tilde{J}}_2$ and $\mathbf{\tilde{M}}_2$ at surface $\mathcal{S}_2$ are easily obtained even for a fine grid size. The regularized fields $\mathbf{\tilde{F}}_{1}$ and $\mathbf{\tilde{H}}_{1}$ of nanorod 1 at the near field surface $\mathcal{S}_2$ have to be computed for each spatial point separately. For more numerical efficiency, we employ two additional simplifications of the surface integral.

The first simplification is based on the symmetry of the QNMs. With the cylindrical metal nanorod, the radiation of the QNMs is axial-symmetric to the nanorod's longitudinal axis. As a result, we only have to calculate $\mathbf{\tilde{F}}_{1}$ and $\mathbf{\tilde{H}}_{1}$ at a line on the side cylindrical surface, a line on the left circular surface and a line on the right circular surface (see Fig.~\ref{fig:s12_Full_vs_pole} (a) bottom) with specific grid size. Then, we extract surface currents $\tilde{\mathbf{M}}_{2},~\tilde{\mathbf{J}}_{2}$ of nanorod 2 at these lines (see Fig.~\ref{fig:s12_Full_vs_pole} (a) top). The dot products of the regularized QNMs and surface currents, namely $\left(\mathbf{\tilde{M}}_2^{*}\cdot\mathbf{\tilde{H}}_1\right)$ and $\left(\mathbf{\tilde{J}}_2^{*}\cdot\mathbf{\tilde{F}}_1\right)$, still have the same axial-symmetry to the longitudinal axis. The final results over the whole surface $\mathcal{S}_{2}$ are obtained by multiplying the results over those three lines by the circumference/(grid size). The integral over $\mathcal{S}_1$ is treated similarly. Note that a similar simplification is also used in Ref.~\onlinecite{ren2020near}.

The second simplification is the use of different grid sizes for the regularized fields and surface currents. For sufficient separation between the rods, the fields of nanorod 1 $\mathbf{\tilde{F}}_{1}$ and $\mathbf{\tilde{H}}_{1}$ do not vary rapidly spatially at the surface $\mathcal{S}_2$ near nanorod 2. Thus, we use a coarse grid size of $5~$nm (see Fig.~\ref{fig:s12_Full_vs_pole} (a) bottom). The surface currents of nanorod 2, $\mathbf{\tilde{J}}_{2}$ and $\mathbf{\tilde{M}}_{2}$ at $\mathcal{S}_2$, meanwhile, change quickly, so we use a grid size of $0.5~$nm (see Fig.~\ref{fig:s12_Full_vs_pole} (a) top). Then, for the dot products of the regularized QNMs and surface currents, we interpolate the regularized QNMs from the coarse grid to the fine grid.

The full frequency integration requires the calculation of the surface integral for each frequency point. With 94 frequency points, the total run time is around $120~$h (for a single MATLAB file in a high-performance computer). For the pole approximation, the run time is around $76~$mins (also for a single MATLAB file). Thus, because of the shorter run time and excellent agreement [cf.~Fig.~\ref{fig:s12_Full_vs_pole}(b)], we
compute $S_{12}^{\rm inter}$ as shown in Fig.~\ref{fig:overlapquant} (b) using the pole approximation from Eq.~\eqref{eq:Sinterdef_pole}.

\red{\section{Calculation details for the quantum Purcell factor}\label{appsec:Purcell}
The dipole coupling Hamiltonian between a two level system (TLS) and the electric field in the rotating wave approximation reads
\begin{align}
    H_{\rm coup} = &-\hat{\sigma}_a^+\int_0^{\infty}\mathrm{d}\omega \mathbf{d}_a\cdot\opvec{E}(\mathbf{r}_a,\omega)\nonumber\\
    &-\hat{\sigma}_a^-\int_0^{\infty}\mathrm{d}\omega \mathbf{d}^*_a\cdot\opvec{E}^{\dagger}(\mathbf{r}_a,\omega)
\end{align}
where \(\hat{\sigma}_a^+\)(\(\hat{\sigma}_a^-\)) creates (annihilates) an excitation in the TLS, and \(\mathbf{d}_a\) is the TLS dipole moment. For a TLS close to one of the rods (e.g. rod 1), the electric field is dominated by the bound QNMs of that rod with only small corrections due to the incoming fields from the second rod (cf.~Appendix~\ref{appsec:incfields}). Using the expansion from Eq.~\eqref{appeq:qnmincprop} as a first approximation of the incoming fields, and employing the residue theorem to solve the frequency integral, yields
\begin{align}
    &\int_0^{\infty}\mathrm{d}\omega\opvec{E}(\mathbf{r}_a,\omega)\Big|_{r_a\in V_1} \approx i\sqrt{\frac{\hbar\omega_1}{2\epsilon_0}}\sqrt{S_{11}}\qnm{f}{1}(\mathbf{r}_a)\hat{a}_1 \nonumber\\
    &\qquad\qquad\qquad+i\sqrt{\frac{\hbar\omega_2}{2\epsilon_0}}\sqrt{S_{22}}\qnm{F}{2}(\mathbf{r}_a,\Tilde{\omega}^*_2)\frac{S^{\rm nrad}_{22}}{S_{22}}\hat{a}_2,
\end{align}
where the first term on the right-hand side represents the bound QNM field of rod 1, while the second term represents the QNM contribution to the incoming fields from rod 2. Furthermore, \(S_{ij} = \delta_{ij}S^{\rm intra}_{ii}+(1-\delta_{ij})S^{\rm inter}_{ij}\) is the overlap matrix.

We next introduce QNM-TLS coupling elements for a TLS near rod 1,
with contributions from both resonator modes
\begin{align}
    g^a_1 &= \sqrt{\frac{\omega_1}{2\hbar\epsilon_0}}\mathbf{d}_a\cdot\qnm{f}{1}(\mathbf{r}_a),\\
    g^a_2 &= \sqrt{\frac{\omega_2}{2\hbar\epsilon_0}}\mathbf{d}_a\cdot\qnm{F}{2}(\mathbf{r}_a,\Tilde{\omega}^*_2)\frac{S^{\rm nrad}_{22}}{S_{22}},
\end{align}
to rewrite the coupling Hamiltonian as
\begin{align}
    H_{\rm coup} = -i\hbar\sum_i \sqrt{S_{ii}}g^a_i\hat{\sigma}_a^+\hat{a}_i +\mathrm{H.a.}
\end{align}

Subsequently,
 to obtain the cavity enhanced spontaneous emission rate, we derive a master equation for the reduced density matrix of the TLS \(\rho_a ={\rm tr}_{\rm em}[\rho]\), where the trace goes over the electromagnetic degrees of freedom, mainly the QNMs. From a second-order Nakajima-Zwanzig equation \cite{breuer2002theory}, we obtain (with operators in the interaction picture)
\sh{Could probably start with the time-local form, and it is not even clear if the non-local form is any more accurate to second order at least, but small point:}
\begin{align}
    &\partial_t \rho_a(t) \nonumber\\
    &\qquad= \int_{t_0}^t\mathrm{d}t' {\rm tr}_{\rm em}\Big\{\big[H_{\rm coup}(t),\big[H_{\rm coup}(t'), \rho_a(t')\otimes\rho_{\rm em}\big]\big]\Big\},
\end{align}
where \(\rho_{\rm em}\) is the initial reduced density matrix of the electromagnetic field, which we assume to be in a vacuum state. We then perform a Born-Markov approximation \cite{breuer2002theory, franke2019quantization, franke2020quantized} and use the commutator \([\hat{a}_i,\hat{a}^{\dagger}_j] = S_{ij}/\sqrt{S_{ii}S_{jj}}\) [cf.~Eq.~\eqref{eq:opsymm}] to obtain a time-local master equation (in the Schrödinger picture) for the reduced density matrix of the TLS:
\begin{align}
    \partial_t \rho_a &= -\frac{i}{\hbar}[\Tilde{H}_a, \rho_a] 
    \nonumber \\
    &+\frac{\Gamma}{2}(2\hat{\sigma}_a^-\rho_a\hat{\sigma}_a^+-\hat{\sigma}_a^+\hat{\sigma}_a^-\rho_a -\rho_a\hat{\sigma}_a^+\hat{\sigma}_a^-),
\end{align}
where \(\Tilde{H}_a\) is the Hamiltonian of just the TLS alone including a photonic Lamb shift \cite{franke2019quantization}, and 
\begin{align} \label{appeq:spemcav}
    &\Gamma = \Gamma_1+\Gamma_2+\Gamma_{12}+\Gamma_{21},\nonumber\\
    &\Gamma_1 = \frac{\red{2}\gamma_1}{\Delta_{1a}^2+\gamma_1^2}S_{11}|g_1^a|^2,\nonumber\\
    &\Gamma_2 = \frac{\red{2}\gamma_2}{\Delta_{2a}^2+\gamma_2^2}S_{22}|g_2^a|^2,\nonumber\\
    &\Gamma_{12} = \mathrm{Re}\Bigg[2g_1^a g_2^{a*}S_{12}\left(\frac{-i\Delta_{2a}+\gamma_2}{\Delta_{2a}^2+\gamma_2^2}\right)\Bigg],\nonumber\\
    &\Gamma_{21}=\mathrm{Re}\Bigg[2g_2^a g_1^{a*}S_{21}\left(\frac{-i\Delta_{1a}+\gamma_1}{\Delta_{1a}^2+\gamma_1^2}\right)\Bigg],
\end{align}
are the cavity-enhanced decay rates due to the modes of rod 1, the modes of rod 2, and the coupling between the modes of the two rods. Here, \(\gamma_i\) is the QNM decay rate, and \(\Delta_{ia} = \omega_i-\omega_a\) is the detuning between the QNM resonance frequency and TLS transition frequency.
As discussed in the main text, the overlap between the modes of far-away cavities is small so that, for sufficiently separated rods, only \(\Gamma_1\) contributes significantly to the overall decay rate of the TLS near rod 1. Thus, for well-separated rods, the single-cavity result from \cite{franke2019quantization} is obtained.
The spontaneous emission rate in a lossless, homogeneous medium is given below Eq.~\eqref{eq:Gamma_QNM}.

}

\bibliography{thebiblio}

\end{document}